 \documentclass[final,3p,times,twocolumn]{elsarticle}

\usepackage{amssymb}
\usepackage{amsmath}
\usepackage{cite}
\usepackage{siunitx}
\usepackage{bm}

\journal{International Journal of Multiphase Flow}

\begin{document}

\begin{frontmatter}



\title{Thin-film flows of granular suspensions on a solid surface}

\author[paris,chicago]{Alice Pelosse} 
\author[paris]{\'{E}lisabeth Guazzelli} 
\author[paris]{Matthieu Roch\'{e}} 

\affiliation[paris]{organization={Université Paris Cité, CNRS, Laboratoire Matière et Systèmes Complexes UMR 7057},
            addressline={10 rue Alice Domon et Léonie Duquet}, 
            city={Paris},
            postcode={75013}, 
            country={France}}
\affiliation[chicago]{organization={James Franck Institute, University of Chicago},
            addressline={929 E 57th St}, 
            city={Chicago},
            postcode={60637}, 
            state={IL},
            country={U.S.A.}}

\begin{abstract}
This review article examines the complex dynamics of thin-film flows of granular suspensions spreading over rigid solid substrates with free air interfaces.
Such systems feature an involved coupling of the free-surface dynamics with the flow and microstructure of the suspension.
In particular, we develop two canonical thin-film situations: drop spreading and dip-coating.
In drop spreading, confinement of the particulate phase near the advancing contact line alters both the spreading rate and the interface shape.
In dip-coating, understanding the entrainment of fluid and particles becomes challenging as the film thickness approaches the particle size.

\end{abstract}

\begin{graphicalabstract}
\end{graphicalabstract}

\begin{highlights}
\item Research highlight 1
\item Research highlight 2
\end{highlights}


\end{frontmatter}

\section{Introduction}
\label{sec:intro}

This review article addresses the free-surface flow of thin films of granular suspensions on rigid solid surfaces, i.e.\,films whose thickness is significantly smaller than their width and length.
The stress balance driving these flows depends on the physical context and scale, see e.g.\,Chapter 4 of \citep{ockendonviscous1995}.
For instance, gravity dominates the spreading of molten lava during volcanic eruptions or the motion of raindrops sliding down a windowpane. 
Conversely, at smaller scales, such as in 3D printing and coating processes, surface tension plays a crucial role in governing flow behavior. 
These systems are inherently multiscale, spanning length scales from the molecular level up to millimeters, and involve the interplay of gravity, capillarity, and viscous forces.

The complexity increases significantly when the flowing fluid is a suspension of solid athermal particles. 
In such granular suspensions, the particle size introduces an additional length scale that must be incorporated into the multiscale framework. 
This is relevant to many practical examples: lava, again, often contains crystallized particles, raindrops may carry particulate matter, and paints or coatings may consist of concentrated suspensions.
Understanding how these particles influence thin-film flow dynamics is thus essential for fundamental scientific research and practical applications.

This review examines how the presence of particles in the liquid, their size and their volume fraction, influence the dynamics of canonical wetting problems, such as droplet spreading or dip-coating \citep{voinov1976hydrodynamics,lopez1976spreading,tanner1979spreading,cox1986dynamics,landau1988,derjaguin1993a}.
Our analysis reveals that while classical continuum models can often be adapted by using a suspension effective viscosity, they break down under conditions of strong particle confinement, where discrete effects become dominant.

This review is intended to provide a didactic introduction to the subject for the multiphase flow community, adopting an experimental perspective.
The structure of the review is as follows.
\S\,\ref{sec:thinFilmFlows} introduces the fundamental concepts of thin-film flows for simple viscous fluids, including droplet spreading and dip-coating. 
\S\,\ref{sec:rheoSuspensions} reviews key aspects of suspension rheology relevant to the granular regime, and focuses on their validity under strong confinement of the particles.
\S\,\ref{sec:granuThinFilms} synthesizes experimental data on thin-film flows of suspensions and interprets these findings in the context of the models discussed in \S\,\ref{sec:thinFilmFlows}. 
Finally, we discuss the effects of inertia or fluid evaporation in such systems, and conclude by outlining promising directions for future research in the study of these flows in \S\,\ref{sec:conclusion}.

\section{Thin-film flows of simple fluids on a solid}
\label{sec:thinFilmFlows}

\S\,\ref{sec:drop-spreading} presents the theory of drop spreading.
We focus on a hydrodynamic, macroscopic description and leave aside microscopic theories \citep{sedev2015}.
Drop profile equations and spreading laws are derived for regular (continuous) fluids. 
\S\,\ref{sec:dip-coating} presents the theory of dip-coating geometry. The law relating the steady-state film thickness to the withdrawal velocity is presented. We also discuss other features of the flow that are important with respect to suspensions.

\subsection{Droplet spreading: general equations}
\label{sec:drop-spreading}

\begin{figure}
    \centering
    \includegraphics[width=0.95\linewidth]{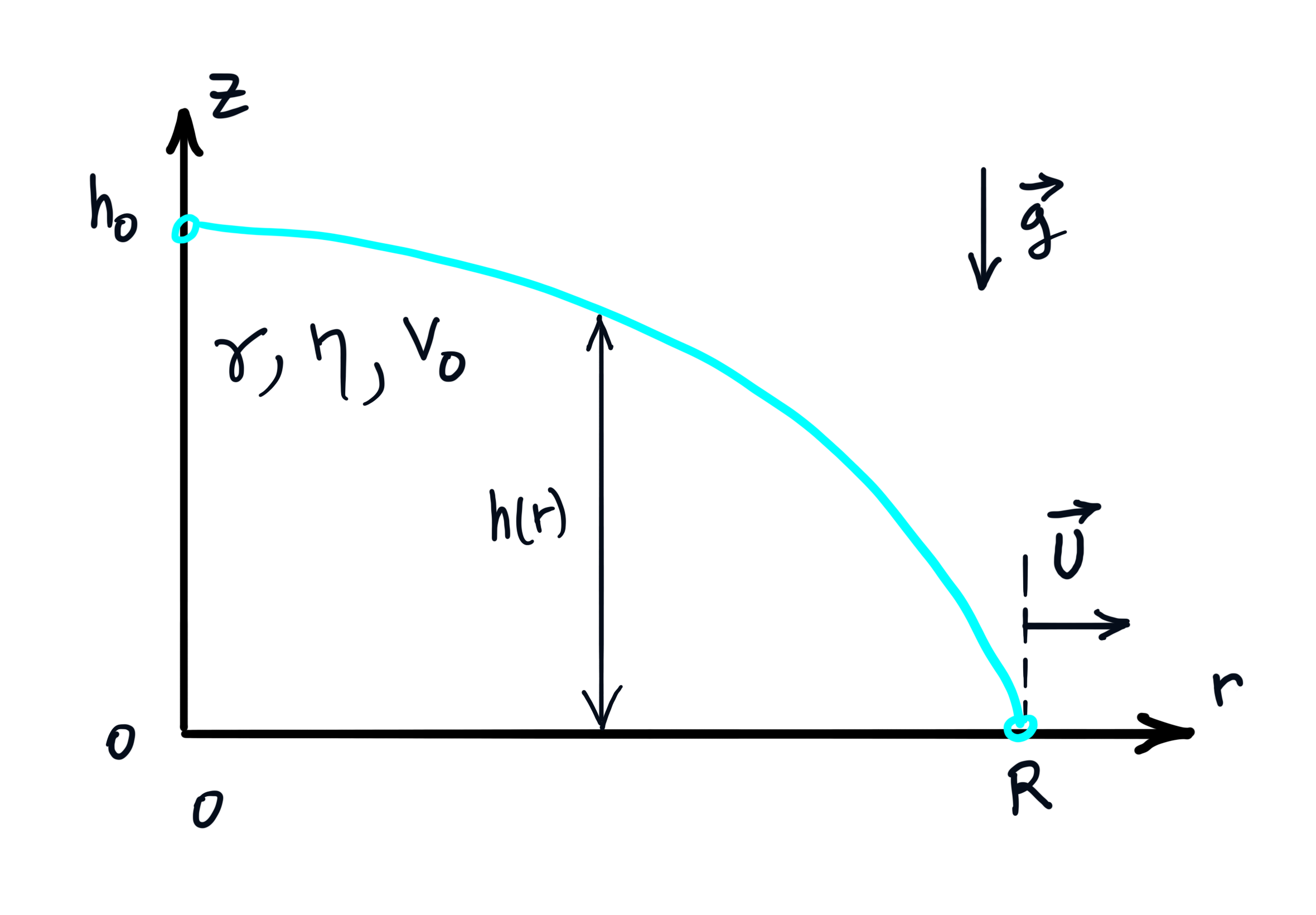}
    \caption{Drop spreading on a solid surface.}
    \label{fig:fig1}
\end{figure}

The spreading of liquid droplets has been the subject of extensive investigation for decades, reflecting both its fundamental interest and its relevance in biological and industrial processes.
In this section, we provide a concise introduction to the problem and the main results needed in \S\,\ref{sec:granuspreading}, to understand what happens when going from a pure fluid to a granular suspension.
Comprehensive reviews can be found in \citep{degennes1985a,bonn2009a}.

We restrict this study to the situation of fluid in complete wetting on a smooth planar solid substrate.
The dynamics of droplet spreading arise from the competition between three fundamental forces: viscosity, capillarity, and gravity.
The resulting differential equation for the profile $h$ of a spreading droplet is \citep{hocking1981,hocking1983}:
\begin{equation}
3\eta U \partial_r h + \frac{1}{r} \partial_r \!\left[ h^2 (h + 3\lambda) r \partial_r \!\left( \gamma (\partial_r^2 h + \frac{1}{r} \partial_r h) - \rho g h \right) \right] = 0,
\label{eq:spreadingf-ED-profile3D}
\end{equation}
with the different quantities involved in Eq.\,\eqref{eq:spreadingf-ED-profile3D} shown in Fig.\,\ref{fig:fig1}. 
We define $r$ as the radial coordinate with $r = 0$ at the center of the drop and $h(r)$ as the axisymmetric drop profile.
The contact line, i.e., the line along which the solid, liquid and gas phases meet, lies at position $r=R(t)$. 
The length scale $\lambda$ is a microscopic slip length that is introduced to circumvent viscous stress divergence at the contact line \citep{navier1823memoire}.
Its order of magnitude is comparable to a molecular size, i.e., a few nanometers.
The gravitational acceleration is noted $g$, and $\eta$, $\gamma$, and $\rho$ are the viscosity, the surface tension, and the density of the liquid, respectively.
This differential equation is solved assuming that the initial drop volume $V_0$ is conserved, and that the liquid/gas interface has a vanishing height at the contact line, and forms a microscopic contact angle $\theta_\text{m}$ with the liquid/solid interface:
\begin{equation}
\int_{0}^{R}2\pi rh(r,t)\text{d}r={V}_0,\; \left.h\right\rvert_{r\rightarrow R}=0,\;\mbox{and}\;
\left.\partial_r h\right\rvert_{r\rightarrow R}=\theta_\text{m}.
\label{eq:boundary_conditions}
\end{equation}      

\begin{figure}
    \centering
    \includegraphics[width=0.99\linewidth]{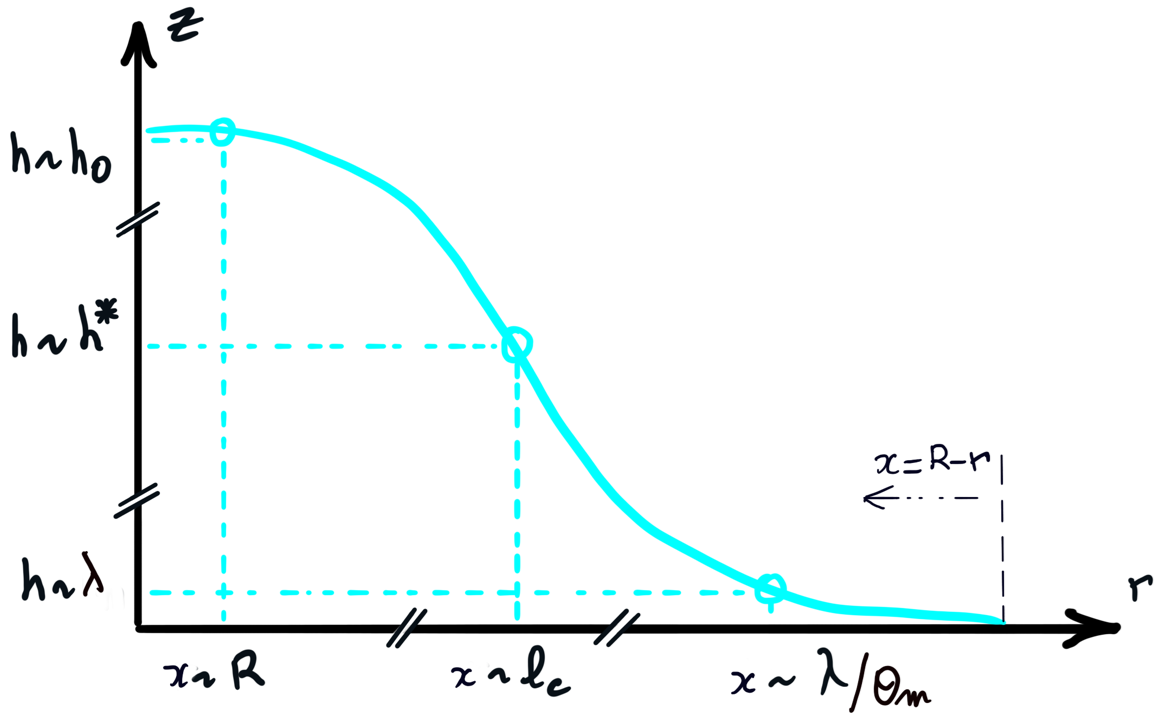}
    \caption{Drop spreading: length scales and regions.}
    \label{fig:fig2}
\end{figure}

Eq.\,\eqref{eq:spreadingf-ED-profile3D} has three intrinsic horizontal length scales depicted in Fig.\,\ref{fig:fig2}.
%
Besides $\lambda$, dimensional analysis indicate that the second scale is the capillary length $\ell_c\gg\lambda$:
\begin{equation}
\ell_c = \left(\frac{\gamma}{\rho g}\right)^{1/2}.
\label{eq:spreadingf-ell-c}
\end{equation}

The physics of objects with sizes smaller (larger) than $\ell_c$ will be dominated by capillarity (gravity).
Finally, the third length scale $R_0 = \left(3V_0/4\pi\right)^{1/3}$ is the typical macroscopic size of the drop.
Only large drops, whose shape is sensitive to gravity, are considered here, $R_0>\ell_c$.
This criterion can be expressed with the Bond number of the problem defined as:
    \begin{equation}
        \text{Bo} = \frac{R_0^2}{\ell_c^2} > 1.
    \label{eq:spreadingf-bond}
    \end{equation}

\subsubsection{Drop profile}
\label{sec:drop-profile}

Because of the large range of scales involved in the description of a spreading droplet, the shape of the interface depends locally on different stress balances.
This feature enables the use of asymptotic matching techniques to address the problem.
While we will not delve into the details of this technique, the description of these regions separately provides key insights into the dynamics of spreading.
We proceed with this description below.

Near the contact line (rightmost region in Fig.\,\ref{fig:fig2}), capillary and viscous forces govern the drop profile, leading to:
\begin{equation}
3\text{Ca} \, \partial_r h + \frac{1}{r} \partial_r \!\left[ h^2 (h + 3\lambda) r \partial_r \!\left( \partial_r^2 h + \frac{1}{r} \partial_r h \right) \right] = 0,
\label{eq:spreadingf-ED-visco-cap}
\end{equation}
with the capillary number, associated with the contact-line motion:
    \begin{equation}
        \text{Ca} = \frac{\eta U}{\gamma}\ll1.
        \label{eq:Ca}
    \end{equation}
This dimensionless number can be interpreted as a normalized velocity, with $\gamma/\eta$ being a viscous-capillary speed characteristic of the interface.
Eq.\,\eqref{eq:spreadingf-ED-visco-cap} is obtained under the quasi-static assumption that the contact line velocity $U$ is small and constant.
The solution of Eq.\,\eqref{eq:spreadingf-ED-visco-cap} must satisfy the boundary conditions:
    \begin{equation}
    \theta(x=\lambda/\theta_\mathrm{m})= \theta_\text{m}\quad \text{and} \quad h(x=\lambda/\theta_\mathrm{m}) = \lambda,
    \end{equation}
with $\theta_\text{m}$ the microscopic contact angle near the contact line, often assumed to be the final equilibrium contact angle of a sessile drop made of the same liquid, and $h_\text{m}$ a microscopic cutoff length scale similar in magnitude to the microscopic slip length $\lambda$.
Eq.\,\eqref{eq:spreadingf-ED-visco-cap} can be transformed into an equation for the apparent dynamic contact angle $\theta_\text{app}$ of the interface.
Its solution is then the Cox-Voinov relation \citep{voinov1976hydrodynamics,cox1986dynamics}:
\begin{equation}
    \theta_\text{app}^3(h)=\theta_\text{m}^3+9\,\text{Ca}\log{\left(\frac{h}{\lambda}\right)}.
    \label{eq:spreadingf-cox-voinov}
\end{equation}
The relation holds for both advancing ($\text{Ca}>0$) and receding ($\text{Ca}<0$) contact lines, and is valid for large slopes, up to $\theta\lesssim3\pi/4$ \citep{voinov1976hydrodynamics,snoeijer2006b}.
A peculiarity of the receding case is that the contact line ceases to exist beyond a velocity threshold \citep{eggers2005}, leading to liquid deposition on the substrate as observed in dip-coating experiments \citep{snoeijer2006c}.

A discussion of the importance of gravity in setting the droplet profile shows that Eq.\,\eqref{eq:spreadingf-cox-voinov} is valid up to a distance of the order of the capillary length $\ell_c$ from the contact line, at which the height of the liquid/gas interface is given by \citep{pelosse2023probing, pelosse2023ecoulements}:
\begin{equation}   
h^{\star}\equiv\ell_{c}\text{Ca}^{1/3}.
    \label{eq:spreadingf-h-star}
\end{equation}
This location corresponds to an inflexion point on the interface.
Relying on this, \citet{pelosse2023probing} tested the relation\,\eqref{eq:spreadingf-h-star} and found good agreement with the experiments.
The typical order of magnitude of $h^{\star}$ across the range of capillary numbers of interest in the study of droplet spreading is 100 \si{\micro\metre}.

\begin{figure}[!ht]
    \centering   \includegraphics[width=0.99\linewidth]{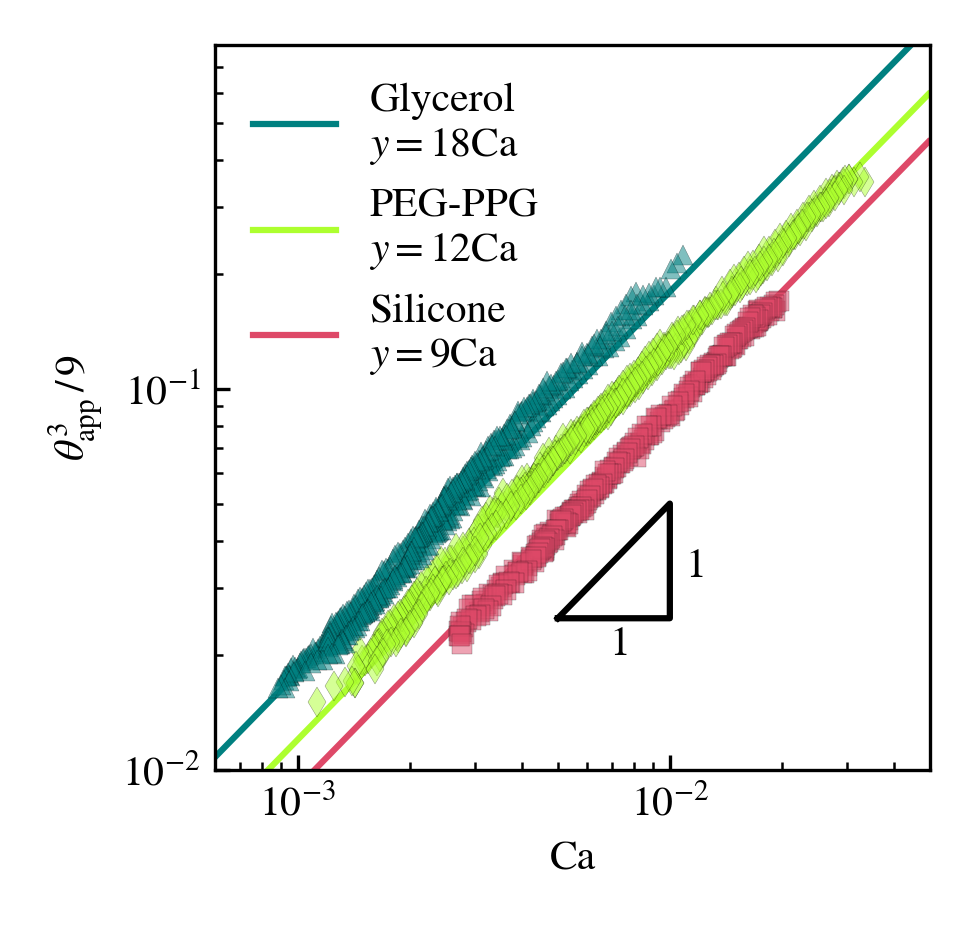}
    \caption{Validation of the Cox-Voinov law adapted from Refs.\,\citep{pelosse2023probing, pelosse2023ecoulements}: $\theta_\text{app}^3/9$ as a function of the capillary number Ca, during the spreading of 300-\si{\micro\liter} drops made of Newtonian fluids: glycerol ($\eta =$\,\SI{1.2}{\pascal\second}, $\gamma=$\,\SI{63}{\milli\newton\per\meter}), poly(ethylene glycol)-ran-poly(propylene glycol) monobutyl ether melt (PEG-PPG, $\eta =$\,\SI{2.3}{\pascal\second}, $\gamma=$\,\SI{35}{\milli\newton\per\meter}), V1000 silicone oil ($\eta =$\,\SI{1.0}{\pascal\second}, $\gamma=$\,\SI{21}{\milli\newton\per\meter}). The fitted value of the factor $\log(x/\lambda)$ is provided in the legend assuming $\theta_\mathrm{m}=0$.}
    \label{fig:CV_regular_fluids}
\end{figure}

Data for the evolution of the dynamic contact angle during droplet spreading as a function of the capillary number for three viscous liquids are displayed in Fig.\,\ref{fig:CV_regular_fluids} \citep{pelosse2023probing, pelosse2023ecoulements}.
The three liquids are perfectly, or quasi perfectly, wetting the solid substrate such that $\theta_\mathrm{m}\sim 0$ and it can be expected from Eq.\,\eqref{eq:spreadingf-cox-voinov} that $\theta_\mathrm{app}^3\propto\mathrm{Ca}$.
As the contact line slows down during spreading, a single experiment sweeps $\text{Ca}$ over more than one order of magnitude.
The apparent contact angle is measured at a drop height, $h =$~\SI{50}{\micro\meter}, well below $h^{\star}$.
For a given fluid, four different experiments are superimposed in Fig.\,\ref{fig:CV_regular_fluids} and show an excellent collapse.
The agreement with the Cox-Voinov law\,\eqref{eq:spreadingf-cox-voinov} is excellent without saturation at low Ca, confirming the hypothesis of quasi-perfect wetting. 
The relative offset of the 3 datasets arises from variations of the
prefactor $\log(x/\lambda)$ as the slip length $\lambda$ changes from one liquid to the other \citep{neto2005boundary, Lauga2007microfluidics}. 

At the macroscopic scale (leftmost region of Fig.\,\ref{fig:fig2}), the shape of the drop is governed by a balance between capillarity and gravity:
\begin{equation}
    \partial_r \!\left[ h^3 r \partial_r \!\left( \partial_r^2 h + \frac{1}{r} \partial_r h - \frac{\rho g}{\gamma} h \right) \right] = 0,
\label{eq:spreadingf-GC-eq}
\end{equation}
with vanishing height at the contact line and volume conservation, reading, respectively:
\begin{equation}
    h(r=R(t)) = 0\quad \text{and}\quad \int_{r=0}^{R(t)} 2\pi r h(r,t) \text{d}r = V_0.
    \label{eq:spreadingf-GC-boundary-conditions}
\end{equation}
For moderate to large Bond numbers, $\text{Bo} \gtrsim 1$, the macroscopic shape of a drop of volume $V_0$ and of contact radius $R(t)$ is given by:
    \begin{equation}
        h(r)=\frac{V_0}{\pi R(t)^2I_2\left(\frac{R(t)}{\ell_c}\right)}\left[I_0\left(\frac{R(t)}{\ell_c}\right)-I_0\left(\frac{r}{\ell_c}\right)\right],
        \label{eq:spreadingf-GC-profile-dim}
\end{equation}
with $I_2$ the  second order modified Bessel function of the first kind \citep{bowman2012introduction}.

\subsubsection{Evolution of the contact radius over time: Tanner's law and derivatives}

Eq.\,\eqref{eq:spreadingf-ED-profile3D} describes the shape of the interface of a spreading drop in general.
Focusing separately on different regions, we obtain local information regarding the shape of the drop.
However, we learn little on the evolution of the outer radius as a function of time, for example.
In what follows, we introduce simple scaling arguments to obtain laws describing spreading at the global, droplet scale.
We focus on two spreading regimes in which viscous dissipation is balanced either by capillarity or gravity.
The former occurs for contact radii $R$ between the droplet and the substrate smaller than the capillary length.
Conversely, the gravity-dominated regime is observed when $R>>\ell_c$.
We also present experimental works found in the literature and compare their results with theory.
%
%

\paragraph{Capillary-driven spreading}


Dimensional analysis can help us establish the equation describing the evolution of contact radius $R$ over time when the contact line moves at velocity $U$.
In the case of small drops, i.e., $\text{Bo}<1$, contact line motion is driven by capillary forces.
We consider a drop profile contained in the $xz$ plane.
For the sake of simplicity, we assume the problem to be invariant in the direction out of the plane, i.e. in the $y$ direction.
Velocity gradients along the vertical coordinate $z$ generate a tangential viscous stress $\eta\partial_z v_x\sim \eta U/h_0$ in the $xz$ plane, $r$ being the spreading direction, $v_x$ the component of velocity along $x$, and $h_0$ the liquid/gas interface height at the center of the drop.
Integrating over $x$, and injecting $U=\text{d}R/\text{d}t=\dot{R}$, we obtain the viscous force per unit length in the $y$-direction $f_v\sim \eta \dot{R} R/h_0$ \citep{cazabat1986dynamics,levinson1988spreading}.
At small contact angles ($\theta_\text{app}\ll 1$) and for a perfectly wetting liquid, the capillary force per unit length in the $y$-direction is $f_c=\gamma_{SG}-\gamma_{SL}-\gamma\cos\theta_\text{app}\simeq \frac{1}{2}\gamma\theta_\text{app}^2$ \citep{de2004capillarity}.
Balancing $f_c$ with $f_v$, we obtain:
\begin{equation}
    \eta \frac{R}{h_0}\dot{R}\sim\gamma
    \theta_\text{app}^2.
\end{equation}

With the approximation $\theta_\text{app}\simeq h_0/R$ and volume conservation, $h_0\sim V_0/R^2$, we end up with: 
\begin{equation}
    \eta R^9\dot{R}\sim\gamma V_0^3\qquad \text{or}\qquad     \frac{\eta}{\gamma}\dot{R}\sim \theta_\text{app}^3,
    \label{eq:tanner-balance-capillarity}
\end{equation} 
which are two equivalent forms of Tanner's law.
The second form can be supplemented with a missing factor to obtain the Cox-Voinov law \eqref{eq:spreadingf-cox-voinov}.
We can also integrate the left expression to predicts radius growth and obtain:
    \begin{equation}
    R(t) = k_\text{c}  \left(\frac{\gamma V_0^3}{\eta}t\right)^{1/10}\propto t^{1/10}.
    \label{eq:tanner-capillarity}
    \end{equation}

\paragraph{Gravity-driven spreading}
Similar reasoning can be carried out with large drops.
In this case, the interface is mostly flat and the driving force is gravity.
The pressure on a vertical column, per unit length in the $y$-direction, is $\int_0^{h_0}\rho g (h_0-z) \,\text{d}z= \frac{1}{2}\rho g h_0^2$, 
that once balanced with the viscous force previously derived leads to:
\begin{equation}
    R^7\dot{R}\sim\frac{\rho g}{\eta}V_0^3.
    \label{eq:tanner-balance-gravity}
\end{equation} 
Equation \eqref{eq:tanner-balance-gravity} therefore yields:
    \begin{equation}
        R(t) = k_\text{g} \left(\frac{\rho g V_0^3}{\eta}t\right)^{1/8}\propto t^{1/8}.
        \label{eq:tanner-gravity}
    \end{equation}

Capillary effects still exist for large drops, especially at the beginning when the drop interface is curved.
Laplace pressure gradients  prevail over gravity at early stages of spreading.
Then, radius growth exhibits two stages: a capillary regime followed by a gravity regime. \citep{cazabat1986dynamics}.
However, once the interface has flattened, gravity becomes the driving force.
The factors $k_\text{c/g}$ in Eqs.\,\eqref{eq:tanner-capillarity} and \eqref{eq:tanner-gravity}, respectively, are \textit{a priori} different.
Few authors comment on these quantities in the literature to the best of our knowledge.
Also, for large drops, a final stage driven by intermolecular forces should exhibit a radius growing like $R\propto t^{1/7}$ \citep{lopez1976spreading,ehrhard1991non}.

\paragraph{Experimental validation}

\begin{table*}
    \centering
    \begin{tabular}{p{0.28\linewidth}p{0.19\linewidth}p{0.11\linewidth}p{0.12\linewidth}p{0.13\linewidth}}
            \textbf{Article}&\textbf{Fluid}& \textbf{Viscosity} & \textbf{Volume} & \textbf{Bond number}\\[0.5ex] 
            \hline\hline
        \textbf{Tanner (1979) \citep{tanner1979spreading} }& Silicone oil & \num{1.08}-\SI{13}{\pascal\second} & \num{0.8}-\SI{3.5}{\micro\liter}&0.2-0.4\\
        \textbf{Cazabat (1986) \citep{cazabat1986dynamics}} & Silicone oil & \num{0.2}-\SI{1}{\pascal\second} & \num{0.35}-\SI{37.9}{\micro\liter}&0.1-2.0\\
        \textbf{Levinson (1988) \citep{levinson1988spreading}}  & Silicone oil & \num{0.2}-\SI{100}{\pascal\second} & \num{0.4}-\SI{40}{\micro\liter}&0.1-2.1\\
        \textbf{Redon (1992) \citep{redon1992spreading}}  & Silicone oil & \SI{11.78}{\pascal\second} & \num{58}-\SI{276.4}{\micro\liter}&2.6-7.4\\
        \textbf{Pelosse (2023 \& 2024) \citep{pelosse2023ecoulements, pelosse2024success}}  & PEG-PPG copolymer & \SI{2.5}{\pascal\second} & \num{100}-\SI{3000}{\micro\liter}&2.2-24\\
        \textbf{Pelosse (2023 \& 2024) \citep{pelosse2023ecoulements, pelosse2024success}}  & Glycerol & \SI{1.2}{\pascal\second} & \num{100}-\SI{1000}{\micro\liter}&1.6-7.7\\
        \hline\hline
\end{tabular}
\caption{Summary of experimental works on drop spreading.}
\label{tab:spreadingt-tanner-exp}      
\end{table*}  

\begin{figure*}[h!]
    \centering
    \includegraphics[width=\linewidth]{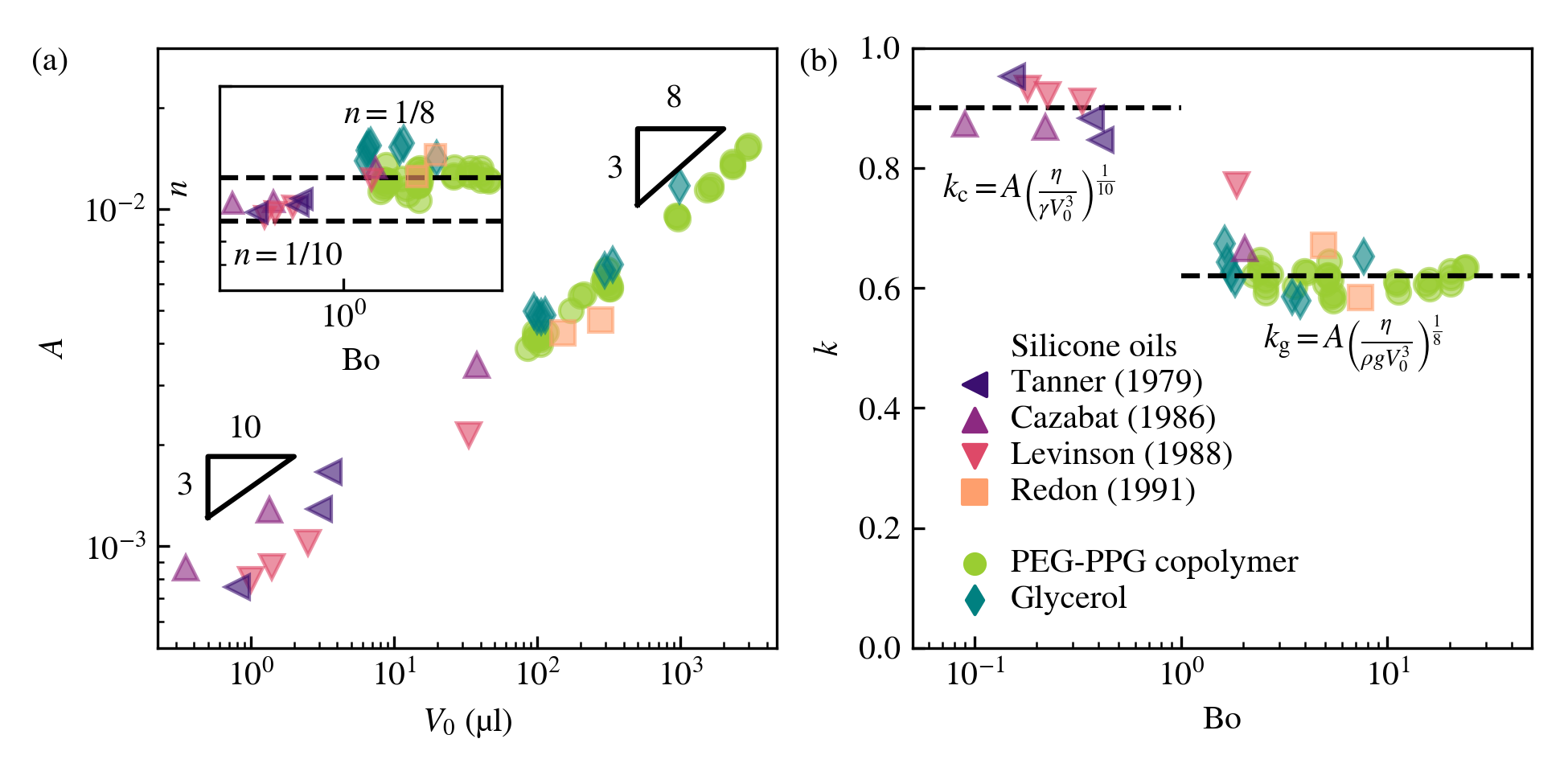}
    \caption{(a) Parameters returned by the fit of experimental results, $R(t) = A(t-t_0)^n$ for drops of viscous fluids. Main graph: factor $A$, inset: exponent $n$. Silicone oil: experimental radius growths extracted from the publications gathered in Table\,\ref{tab:spreadingt-tanner-exp}  \protect\footnotemark[1]. 
    (b) Factor $k_\text{c/g}$ of the spreading law as a function of the drop Bond number, $\text{Bo}=(3V_0/4\pi)^{2/3}/\ell_c^2$, computed according to the spreading regime, i.e., $k_\text{c}=A\left[\eta/(\gamma V_0^3)\right]^{1/10}$ for $\text{Bo}<1$ and $k_\text{g}=A\left[\eta/(\rho g V_0^3)\right]^{1/8}$ for $\text{Bo}>1$. Capillary regime: $k_\text{c}= 0.90\pm = 0.03$, gravity regime: $k_\text{g}= 0.62\pm0.02$.}
    \label{fig:spreadingt-radius-biblio}
\end{figure*}

Table\,\ref{tab:spreadingt-tanner-exp} summarizes the systems used by various authors to test Eqs.\,\ref{eq:tanner-capillarity} and \ref{eq:tanner-gravity}.
Silicone oils are used extensively because of the large range of viscosities available.
Moreover, these oils are insensitive to evaporation and hygroscopic effects.
Two other viscous fluids, a Newtonian copolymer melt (Poly(ethylene glycol-ran-propylene glycol) monobutyl ether, $M_w=3,900$, Sigma) and glycerol, are also used \citep{pelosse2023ecoulements, pelosse2024success}.
In all these experiments, radius growth follows a power law and datasets are fitted with $R(t) = A ( t - t_0)^n$.
The parameter $t_0$ accounts for uncertainties in the starting time of the experiment, and has no physical significance.
First, the data divide into two groups with respect to exponent $n$, see inset of Fig.\,\ref{fig:spreadingt-radius-biblio}(a).
For large Bond numbers, $n$ lies around $1/8$, while it is closer to $1/10$ when $Bo<1$.
%
%
Second, values of prefactor $A$ are plotted in the main plot in Fig.\,\ref{fig:spreadingt-radius-biblio}(a), as a function of drop volume $V_0$.
\footnotetext[1]{ \citet{tanner1979spreading} computes the volume from the drop radius and contact angle of a spherical cap: $V_0=\frac{\pi}{2}R^2h_0\left(1+\frac{3h_0^2}{2R^2}\right)=\frac{\pi R^3}{2}\frac{1-\cos(\theta)}{\sin(\theta)}\left(1+\frac{3(1-\cos(\theta))^2}{2\sin(\theta)^2}\right)$ with $\tan(\theta)=2h_0R/(R^2-h_0^2)$} 
The dataset is compatible with theoretical predictions, $A\propto V_0^{3/10}$ when $Bo<1$ and $V_0^{3/8}$ when $Bo>1$, within experimental uncertainties.
In Fig.\,\ref{fig:spreadingt-radius-biblio}(b), normalization of $A$ by a combination of volume, density and surface tension, and viscosity effects dependent on the spreading regime helps determine the values of of $k_\text{c}=A\left[\eta/(\gamma V_0^3)\right]^{1/10}$ for $\text{Bo}<1$, and $k_\text{g}=A\left[\eta/(\rho g V_0^3)\right]^{1/8}$ for $\text{Bo}>1$ in Eqs.\,\eqref{eq:tanner-capillarity} and \eqref{eq:tanner-gravity}, respectively.
The values collapse on $k_\text{c}= 0.90\pm0.03$ and $k_\text{g}= 0.62\pm0.02$, with little dependence on the nature of the liquid.

Thus, experiments confirm the theoretical predictions of Eqs.\,\eqref{eq:tanner-capillarity} and \eqref{eq:tanner-gravity} regarding drop spreading, and more specifically, the radius power-law growth with a transition from a capillary- to a gravity-driven dynamics around $\text{Bo}=1$, as shown in the inset of Fig.\,\ref{fig:spreadingt-radius-biblio}(a).

\subsection{Dip-coating}
\label{sec:dip-coating}

Another classic configuration to study thin film problems is dip-coating: a vertical (or inclined) plate is withdrawn from a bath at constant speed $U$, as sketched in Fig.\,\ref{fig:dipSchematic}. 
Depending on the withdrawing velocity and the wetting conditions, a liquid film may coat the plate.
We recall below how the steady-state film thickness, $h_0$, depends on experimental parameters.
Note that the plate may also be forced into the bath.
This topic will be left aside here.
The presentation below is inspired from references \citep{deryck1998,quere1998,quere1999,rio2017a}.

\subsubsection{Complete wetting}
\label{subsec:compWet}
\citet{landau1988} and \citet{derjaguin1993a} (referred to as LLD in the following) separately proposed a solution to the dip-coating problem that relates $h_0$ to $U$ when the liquid completely wets the solid. 
The key insight here is to realize that the problem can be divided into three regions ruled by different physics, making the system amenable to asymptotic matching techniques.
Here, we recall the main theoretical arguments and comment on a few other aspects of this problem that are important to understand dip-coating with granular suspensions addressed in \S\,\ref{sec:granudipcoating}.

\begin{figure}[!ht]
    \centering
    \includegraphics[width = \linewidth]{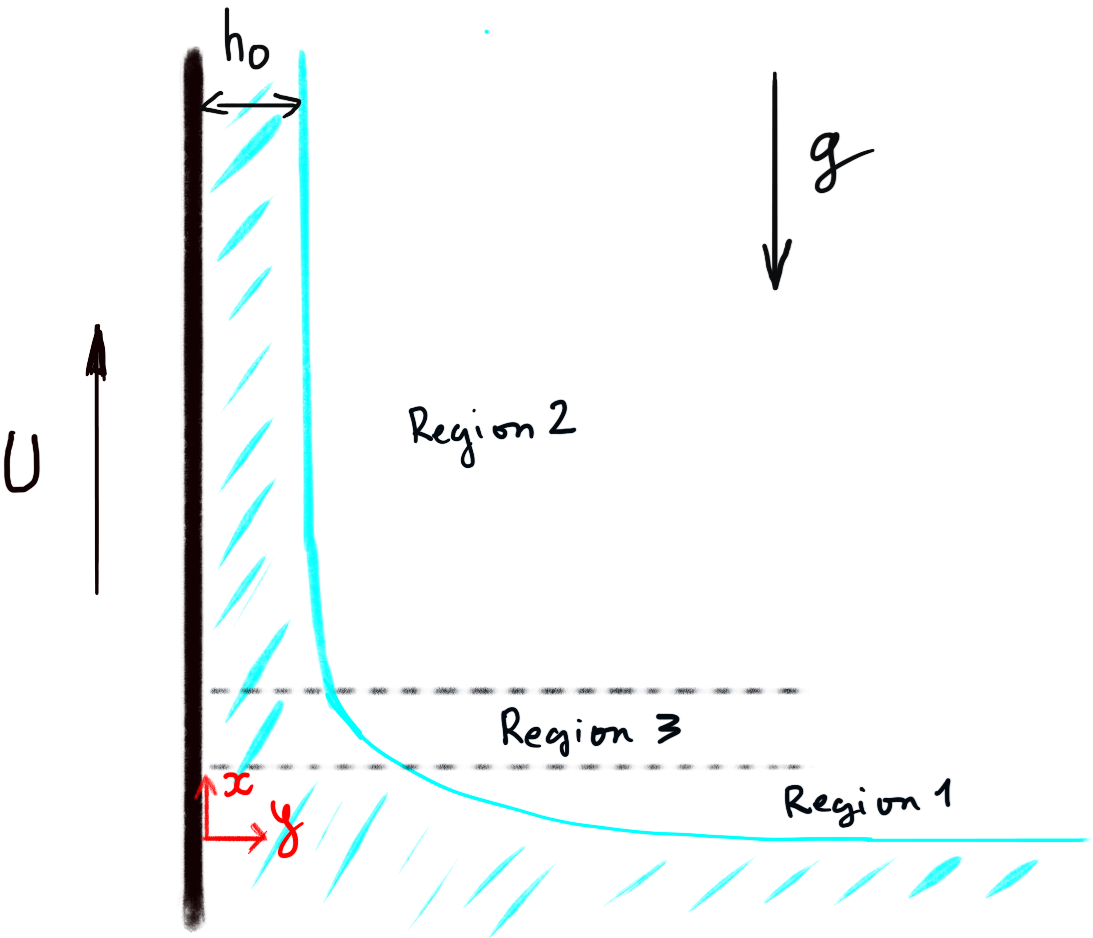}
    \caption{Sketch of the dip-coating experiment.}
    \label{fig:dipSchematic}
\end{figure}
Figure \ref{fig:dipSchematic} depicts the problem.
The plate is immersed in the liquid before the experiment starts.
The liquid wets and rises on the plate. 
A static meniscus (region 1) forms that has a curvature $\kappa_\text{sm}=\sqrt{2}{\ell_c}^{-1}$ \citep{de2004capillarity}.
\citet{landau1988} base their model on the fact that setting the plate in motion has two consequences. 
First, a thin film of liquid with thickness $h_0$ (region 2) is entrained. 
Second, the meniscus deforms to accommodate viscous stresses.
If the plate velocity $U$ is small enough (but not too small, so that intermolecular forces can be neglected \citep{quere1989}), this deformation is a perturbation to the static meniscus shape, called the dynamic meniscus (region 3).

The theoretical description goes as follows. 
The system is assumed translation-invariant along the third dimension.  The velocity has a single component, $u$, along $\bm{x}$. 
Inertia is neglected and the slopes of the liquid/gas interface are small. 
The contact line on the plate is far above the surface of the bath. 
Finally, the film thickness $h_0$ is constant and much smaller than the film width and length. 
Then, the lubrication equations write: 
\begin{equation}
    \partial_x p = \eta \partial_{yy}u - \rho g\quad \text{and}\quad 
    \partial_y p =0 \label{eq:lub}. 
\end{equation}
Pressure is thus constant along $y$, and it depends on $x$ through the Laplace pressure.
Under the small-slope assumption, the curvature of the liquid-gas interface is given by $\kappa\simeq\partial_{xx}h$, and $p=p_0-\gamma\kappa$, where $p_0$ is the ambient pressure and $\gamma$ is the surface tension of the liquid.

Eqs.\,\eqref{eq:lub} describe the three regions delineated by \citet{landau1988}. 
For example, the solution to these equations when $u=0$ describes the shape of the static meniscus.
Moreover, in the flat film, the Laplace pressure term vanishes, and the thickness of the film is the solution of the reduced equation $\eta\partial_{yy}u=\rho g$.
However, the full equations must be solved to find the shape of the dynamic meniscus.

In what follows, we focus on the dynamic meniscus. 
Complementing Eqs.\,\eqref{eq:lub}  with a no-slip boundary condition at the surface of the substrate and continuity of stresses at the liquid-gas interface, and integrating, the expression of the velocity field writes:
\begin{equation}
    u(y)=\left(\frac{\rho g - \gamma\partial_{xxx}h}{\eta}\right)\left(\frac{y^2}{2}-hy\right)+U.
    \label{eq:LLD_u}
\end{equation}

The next step is to express the mass flow rate (per unit length) in the dynamic meniscus using Eq.\,\eqref{eq:LLD_u}.
The following expression is obtained:
\begin{equation}
    Q_\text{dm}=\int_0^h u~dy= \frac{\gamma\partial_{xxx}h-\rho g}{3\eta}h^3+Uh.
    \label{eq:flowRate_dm}
\end{equation}
This expression can be used to obtain the flow rate in the flat thin film by setting the capillary term to 0:
\begin{equation}
    Q_\text{tf}=Uh_0-\frac{\rho g}{3\eta}{h_0}^3.
    \label{eq:flowRate_tf}
\end{equation}
Mass conservation imposes that $Q_\text{dm}=Q_\text{tf}$, and we obtain:
\begin{equation}
    \partial_{xxx}h=\frac{1}{\ell_c^2}\frac{h^3-{h_0}^3}{h^3}+3\text{Ca}\frac{h_0-h}{h^3}.
    \label{eq:sol1}
\end{equation}
This equation can be made dimensionless using $x=h_0\text{Ca}^{-1/3}X$, and $h=h_0H$, leading to:
\begin{equation}
    \partial_{XXX}H=3\frac{1-H}{H^3}+\frac{{h_0}^2}{\text{Ca}~{\ell_c}^2}\left(\frac{H^3-1}{H}\right).
    \label{eq:sol2}
\end{equation}
The second term of the r.h.s.\,accounts for the contribution of gravity.
For the sake of simplicity, we assume that $(h_0/\ell_c)^2<< \mathrm{Ca}$.
Eq.\,\eqref{eq:sol2} then reads:
\begin{equation}
    \partial_{XXX}H=3\frac{1-H}{H^3}.
    \label{eq:LLDeq}
\end{equation}

The conceptual breakthrough was to realize that considerations based on matched asymptotic expansions could lead to a relation between film thickness $h_0$ and withdrawal velocity $U$.
The dynamic meniscus branches both into the thin film and in the static meniscus.
The latter imposes that the curvature of the interface of the dynamic meniscus be equal to that of the static meniscus at the branching point.
Labeling dimensionless coordinates in the static (dynamic) meniscus as $H_\text{sm}$ ($H_\text{dm}$), this constraint leads to \citep{landau1988}:
\begin{equation}
    \partial_{X_\text{dm}X_\text{dm}}H_\text{dm}\Big\rvert_{H_\text{dm}\rightarrow\infty}=\partial_{X_\text{sm}X_\text{sm}}H_\text{sm}\Big\rvert_{H_\text{sm}\rightarrow 0}.
    \label{eq:curvMatch}
\end{equation}
Given $\kappa_\text{sm}=\sqrt{2}\ell_c^{-1}$, and our choice for non-dimensionalization, we find that:
\begin{equation}
    \partial_{X_\text{dm}X_\text{dm}}H_\text{dm}\Big\rvert_{H_\text{dm}\rightarrow \infty}=\frac{h_0}{\text{Ca}^{2/3}}\kappa_\text{sm}.
    \label{eq:curvMatch2}
\end{equation}
Numerical integration of Eq.\,\eqref{eq:LLDeq} shows that $\partial_{X_\text{dm}X_\text{dm}}H_\text{dm}\Big\rvert_{H_\text{dm}\rightarrow\infty}\simeq 1.34$.
Rearranging and moving back to dimensional coordinates, the Landau-Levich-Derjaguin (LLD) law then follows:
\begin{equation}
    h_0=0.95~\text{Ca}^{2/3}~\ell_c.
    \label{eq:LLDlaw}
\end{equation}
The thickness of the thin film coating the substrate is the product of a sublinear power law of the dimensionless withdrawing velocity $\text{Ca}$ and of the capillary length $\ell_c$ of the liquid.

\begin{figure}[!ht]
    \centering
    \includegraphics[scale=.925]{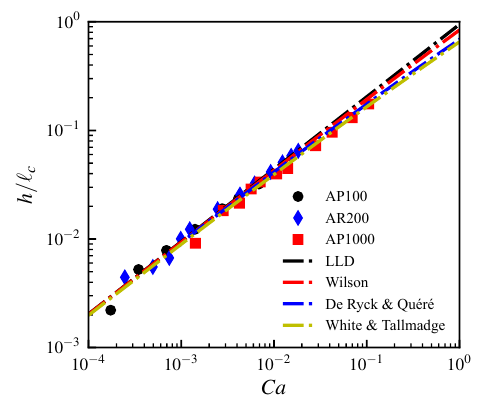}
    \caption{Comparison of the Landau-Levich-Derjaguin (LLD) law \citep{landau1988,derjaguin1993a}, its extensions to larger capillary numbers by \citet{white1965}, \citet{wilson1982}, \citet{deryck1998} to experimental data obtained by \citet{gans2019dip} during the withdrawal of a glass plate from baths of three different silicone oils. } 
    \label{fig:summaryLLD}
\end{figure}
Eq.\,\eqref{eq:LLDlaw} is expected to be valid when $\text{Ca}<10^{-3}$. 
Over the years, solutions to the dip-coating problem at larger capillary numbers have been proposed using different strategies such as an extension of the technique used by Landau and Levich \citep{white1965}, asymptotic matching \citep{wilson1982}, and numerics \citep{deryck1998}.
These solutions all improve the matching with experiments, with the results of \citet{deryck1998} and \citet{white1965} being the closest to experimental data obtained by \citet{gans2019dip}, see Fig.\,\ref{fig:summaryLLD}.
Thus, we will use the White-Tallmadge (WT) correction for the remainder of the review,
\begin{equation}
    \text{Ca}\simeq 1.09 \left(\frac{h_0}{\ell_c}\right)^{3/4}+\frac{h_0}{\ell_c}.
    \label{eq:WT}
\end{equation}

In the context of this review, it is important to note that there exists a stagnation point in dip-coating flows
that divides the flow between a sheared region close to the wall and a recirculating volume away from it \citep{colosqui2013b}.
Using mass conservation, $Q_\text{dm}=Q_\text{tf}$, and replacing the gravity-capillary prefactor in \eqref{eq:LLD_u} before setting this velocity equal to 0 at the point of stagnation for a height $y=h^\star_{LLD}$, we obtain:
\begin{equation}
    h^\star_\text{LLD}=\left(3-\frac{h_0^2}{\text{Ca}~\ell_c^2}\right)h_0.
    \label{eq:stagPos}
\end{equation}
We will come back to its significance in \S\,\ref{sec:granudipcoating} for dip-coating of granular suspensions.

\subsubsection{Partial wetting}
\label{subsec:partWet}

\begin{figure}[!ht]
    \centering
\includegraphics[width=1\linewidth]{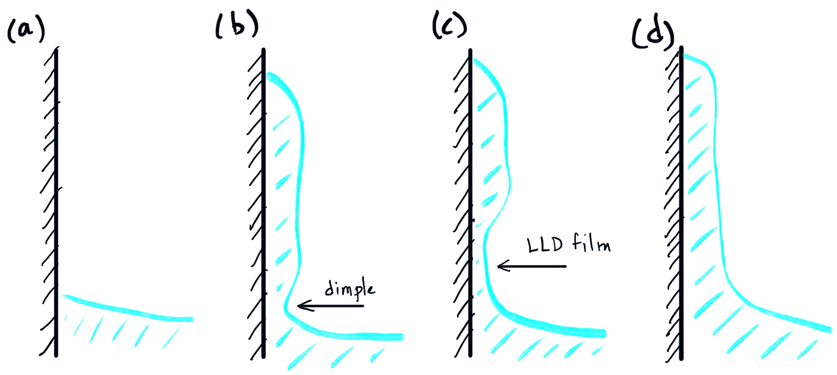}
    \caption{Fluid entrainment by a moving wall in partial wetting. 
    The capillary number increases from (a) to (d).}
    \label{fig:partialwetting}
\end{figure}
In many cases, the liquid wets the solid partially, i.e., the liquid/gas interface forms a finite angle with the solid/liquid interface at the contact line \citep{de2004capillarity}.
The dynamics of dip-coating then becomes more complex.
Perhaps the most spectacular difference is that there is a threshold capillary number $\text{Ca}_\text{e}$ for film entrainment \citep{snoeijer2006c,snoeijer2008a,delon2008,ziegler2009,gao2016e}. 
Below this threshold, the meniscus formed by the liquid rises to higher and higher positions as the velocity is increased, but no liquid is entrained, see Fig.\,\ref{fig:partialwetting}(a).
The contact angle of the receding contact line on the wall is then given by the Cox-Voinov law Eq.\,\eqref{eq:spreadingf-cox-voinov}, with the capillary number being negative.
Beyond $\text{Ca}_\text{e}$, a film much thicker than the LLD prediction\,\eqref{eq:LLDlaw} coats the plate, see Fig.\,\ref{fig:partialwetting}(b).
This film connects to the bath through a dimple.
The height of this dimple decreases as $\text{Ca}$ increases.
Beyond a second capillary threshold, the dimple is replaced by a LLD film described by Eq.\,\eqref{eq:LLDlaw} connected to the thick film by a capillary shock, see Fig.\,\ref{fig:partialwetting}(c).
This LLD film thickens with an increase of $\text{Ca}$, until it has the same thickness as the thick film.
Then, a film of monotonic thickness covers the plate and branches into the bath, see Fig.\,\ref{fig:partialwetting}(d).

It should be noted that the exact value of the threshold capillary number $\text{Ca}_\text{e}$ depends on the shape of the substrate.
While the transition to liquid entrainment by a withdrawn cylindrical fiber occurs when the contact angle at the top of the meniscus vanishes \citep{sedev1991a}, as predicted by theory, the threshold capillary number is smaller on a flat plate, suggesting that the transition occurs before the contact angle nullifies \citep{snoeijer2006c}.
The origin of this discrepancy is still under investigation.

\section{Rheology of granular suspensions under confinement}
\label{sec:rheoSuspensions}

The rheology of granular suspensions depends directly on the microstructure adopted by the particles under flow.
This microstructure is affected by particle confinement, i.e., when the size of the flow and the particle diameter are comparable.
Such a situation occurs also in thin free films, albeit with the difference that particles are confined by a the deformable free surface.
Understanding confinement effects in suspensions is therefore essential for predicting and controlling the dynamics of such films. 
In the following, we first introduce the basis of granular suspension rheology and then focus on how confinement modifies their rheological behavior.
These concepts will be useful to describe better and understand suspension thin-film flows discussed in the following \S\,\ref{sec:granuThinFilms}.

\subsection{Rheology of granular suspension}

Let us define a particle confinement parameter $\xi=h/d$, where $h$ is the characteristic size of the flow and $d$ the particle diameter. At the macroscopic scale, when $\xi>>1$, a suspension behaves as an effective continuous fluid, thus justifying the use of an apparent bulk viscosity $\eta_\text{bulk}$ to quantify its resistance to flow. 
The viscosity of  Newtonian granular suspensions does not depend on particle size and it increases with the particle volume fraction $\phi$ and suspending fluid viscosity $\eta_0$. 
It is actually proportional to the viscosity of the suspending liquid:
\begin{equation}
    \eta_\mathrm{bulk}=f(\phi)\eta_0,
\end{equation}
with the complete form of $f(\phi)$ being only empirically known \citep{GuazzelliPouliquen2018}.
Divergence occurs at the maximum flowable volume fraction $\phi_c$ \citep{GuazzelliPouliquen2018} that depends on the size distribution of the particles, their rigidity, as well as on their hydrodynamic and contact interactions
\citep{ness2022a}.
The balance of the hydrodynamic and contact contributions depends on $\phi$.
In particular, contact interactions prevail over hydrodynamic interactions in the dense regime ($\phi\gtrsim 40\%$) \citep{gallier2014rheology} that is of interest in this review.
As a consequence, the particle microstructure has a major influence on the macroscopic bulk flow \citep{morris2009review}. 

\subsection{Effect of confinement on suspension rheology}

A decrease of the system size down to a few particle diameters, $\xi\rightarrow 1$, such as when blood flows through thin capillaries \citep{fahraeus1931viscosity}, profoundly modifies the picture we have just introduced.
These observations can be rationalized by several factors inducing modifications of the particle spatial distribution under confinement, such as irreversible inhomogeneity in concentration created by shear \citep{gadala1980shear, abbott1991experimental, han1999particle, snook2016dynamics}.
This phenomenon, referred to as shear-induced particle migration, is understood as a particle flux toward regions of low shear, leading to a decrease in the effective viscosity of the confined suspension \citep{leighton1987shear}.

\begin{figure}
    \centering
    \includegraphics[width=0.75\linewidth]{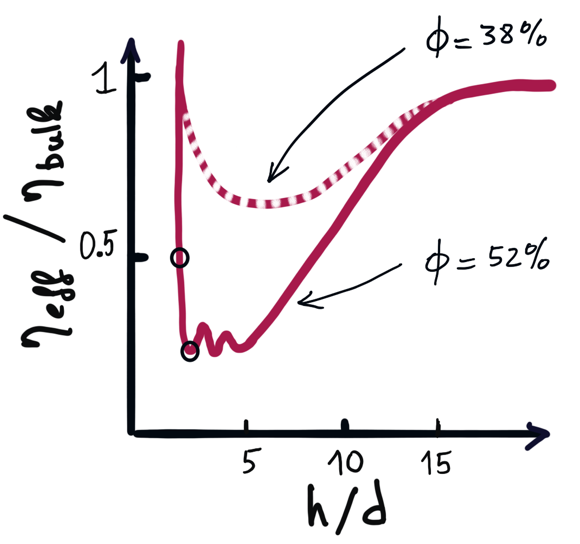}
    \includegraphics[width=0.48\linewidth]{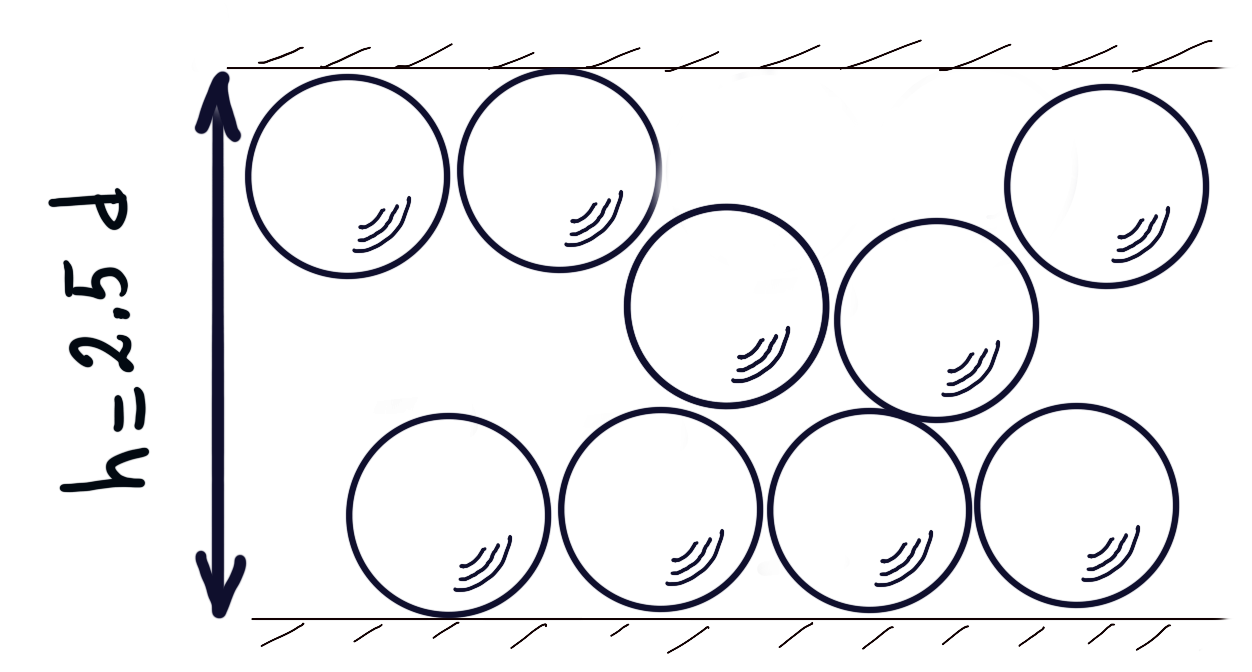}
    \includegraphics[width=0.48\linewidth]{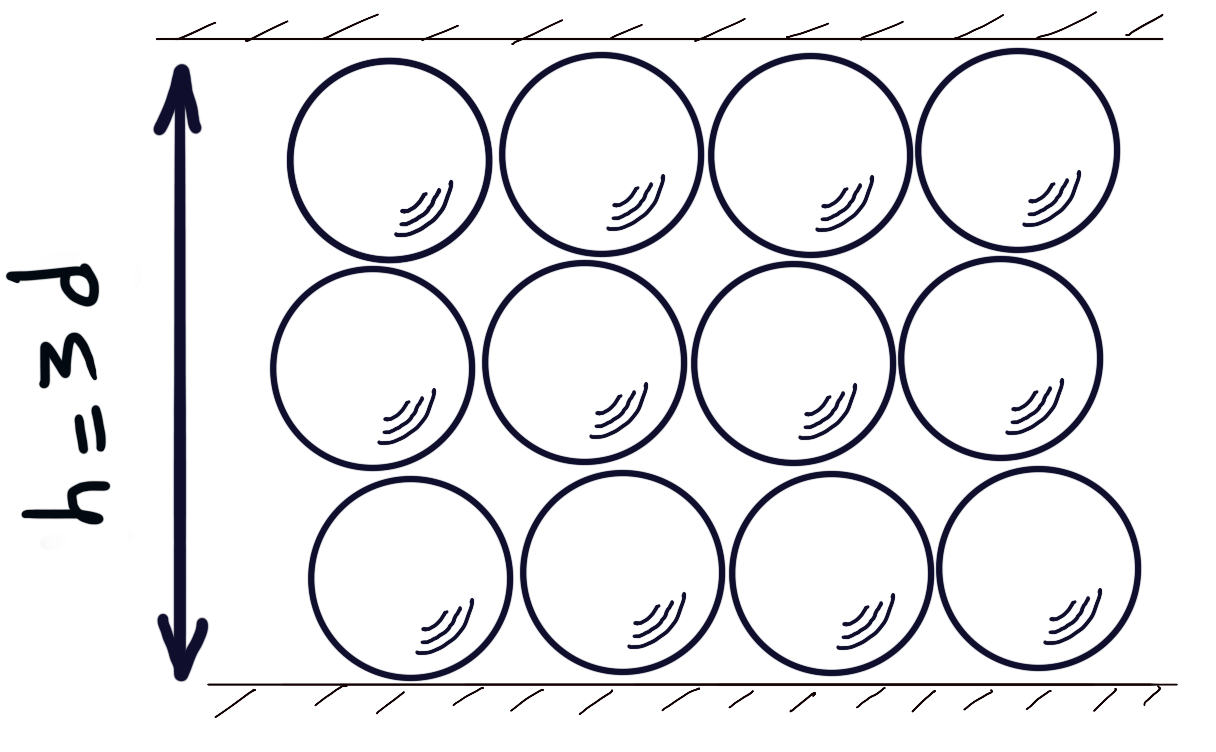}
    \caption{Confinement of a suspension between two rigid walls spaced by a distance $h$. (Top) Effective suspension viscosity normalized by the bulk value as a function of particle confinement $h/d$. (Bottom) Sketches of the situations corresponding to the circles on the top graph. (Left) Gap is not commensurable with the particle diameter. (Right) Gap is a multiple integer of the particle diameter and corresponds to a minimum of viscosity. }
    \label{fig:fig4}
\end{figure}

In addition, when $\xi\sim 1$, the viscosity becomes dependent on the commensurability between $d$ and $h$.
For example, microrheology experiments have shown that under strong confinement, the effective viscosity of the suspension is an oscillating function of the gap size that reaches a local minimum when $\xi$ is a (small) integer, as depicted in Fig.\,\ref{fig:fig4} \citep{ramaswamy2017confinement}. 
For these small integer values of\,$\xi$, the particles organize in stable, low dissipation, layers across the gap.
In this near-wall ordering process, particle layers tend to form a crystalline hexagonal structure at volume fractions significantly lower than those needed for crystallization in the bulk \citep{yeo2010ordering, gallier2016effect}.
Polydispersity disrupts the formation of large-scale microstructures and prevents the viscosity reduction that would otherwise arise from particle ordering.
Changes in $\xi$ away from an integer value lead to defects in the layering and an increase in the effective viscosity of the suspension.
Numerical simulations reproduce these oscillations of $\eta_\mathrm{eff}$ with the local minima when $\xi$ is an integer, as well as the particle ordering \citep{fornari2016rheology}, which is attributed to the creation of a particle-depleted region near the wall that induces layering at distances greater than one particle diameter from the boundary, when $\xi<15$.
This layering spans the entire system in confined situations, especially as the packing fraction increases \citep{gallier2016effect}.

Similar experiments carried out on a regular rheometer with dilute granular suspensions ($\phi\leq0.2$) at moderate confinement, shed another light on the effect of particle confinement \citep{peyla2011new}.
A monotonic increase of the viscosity as $\xi\rightarrow 1$ has been reported.
These contrasting results may stem from the fact that ordering fades rapidly when decreasing the volume fraction \citep{gallier2016effect}.
Hence, the apparent viscosity does not display a decrease nor local minima. 
Moreover, confined flows could affect the balance between hydrodynamic and contact interactions differently depending on the volume fraction.
In 2D simulations, dilute and semi-dilute suspensions exhibit an increase of hydrodynamic interactions (and thus of the effective viscosity) under confinement because of an enhancement of dissipation near the wall \citep{doyeux2016effective}.

However, when two particles align along the shear direction, a dip in the dissipation of energy would appear in-between, explained by the screening of the imposed shear by the particles. 
This anomaly can lead to a strong decrease in the overall dissipation, but depends on the proportion of aligned particles in the confined shear flow.

\begin{figure*}[!htb] 
    \centering
    \includegraphics[width =  \linewidth]{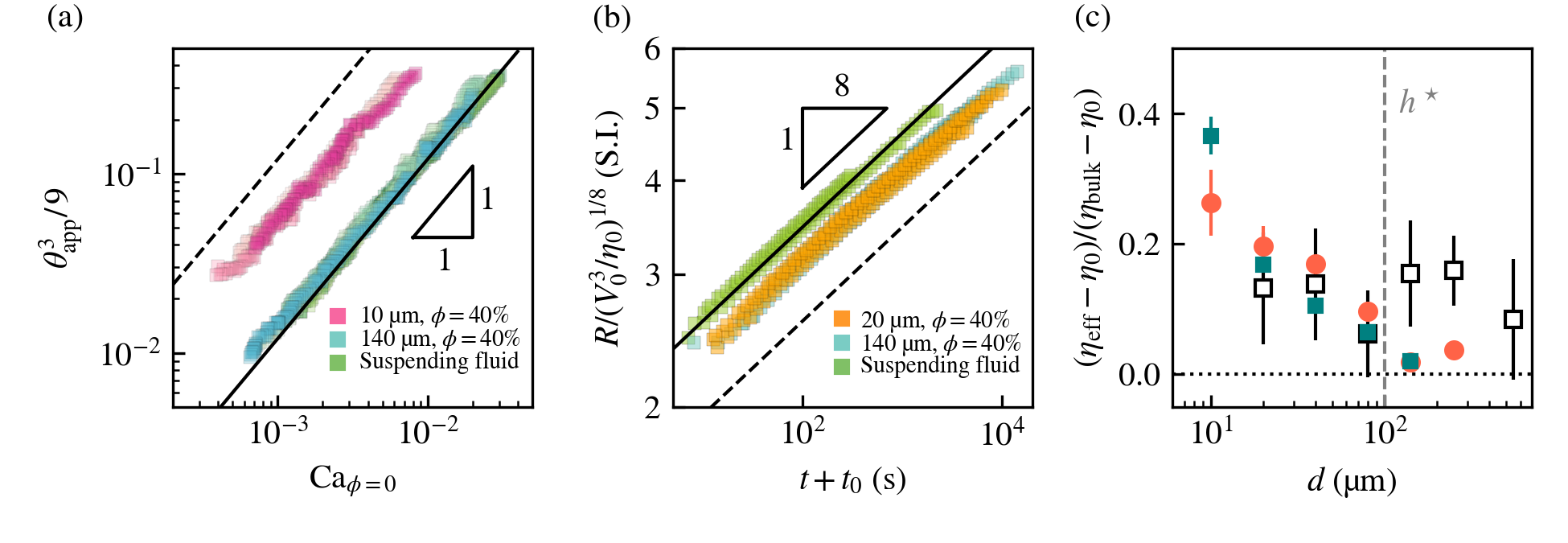}
    \caption{Validation of the Cox-Voinov law and radius power-law adapted from Ref.\,\citep{pelosse2023ecoulements} for drops made of pure PEG-PPG and monomodal suspensions with different PS particles (see legend). (a) Cube of the apparent contact angle, $\theta_\text{app}^3/9$, as a function of $\mathrm{Ca}_{\phi=0}$, the capillary number of the suspending fluid (PEG-PPG), during the spreading of 300-\si{\micro\liter} drops. 
    (b) Radius normalized by volume and viscosity effect as a function of time, during the spreading of 100-\si{\micro\liter} drops. 
    Solid lines: best fitted slope for the pure fluid alone. 
    Dashed-lines: prediction extrapolated from the suspension bulk viscosity. (c) Difference between the wetting viscosity $\eta_{eff}$ and the viscosity of the suspending liquid $\eta_0$ rescaled by the difference between the bulk viscosity $\eta_{bulk}$ and $\eta_0$ inferred from (solid symbols) the Cox-Voinov law \eqref{eq:spreadingf-cox-voinov} at $\phi = 30\%$ (orange circles) and $\phi = 40\%$ (blue squares) and from the gravity-viscous spreading law \eqref{eq:tanner-gravity} (open symbols)  at $\phi = 40\%$ as a function of particle size $d$. This quantity is equal to 0 when $\eta_{\rm eff}=\eta_{0}$ and 1 when $\eta_{\rm eff}=\eta_{\rm bulk}$. Adapted from \citep{zhao2020spreading, pelosse2023ecoulements, pelosse2024success}. 
    The height of the viscous cut-off $h^\star$ is indicated by the vertical gray dashed line.}
    \label{fig:CV_Tanner_susp}
\end{figure*}

In conclusion, confinement alters the suspension microstructure and macroscopic dissipation. 
Experiments and numerical investigations indicate that its effects appear for flow sizes smaller than 20 particle diameters. 
However, its global effect on suspension rheology depends on the particle volume fraction. 
Confinement can locally decrease particle concentration through shear-induced particle migration and trigger particle layering near the wall. 
Under extreme confinement, commensurability of the system size with a particle diameter could be critical in the overall dissipation and can be rationalized by the stability of the layers of particles. 
Also, while the hydrodynamic contribution to dissipation would increase with confinement, dense systems could be less dissipative because of shear screening by the particles. 
Addressing these questions remains challenging both experimentally and numerically and requires further study.

\section{Thin-film flows of granular suspensions}
\label{sec:granuThinFilms}

We now turn to thin-film flows of granular suspensions.
These systems challenge the standard effective-medium description of suspensions as the size of the flow compares to that of the particles.
By contrast with the results presented in the previous section, particle confinement is imposed by a deformable, liquid/gas, free interface.

\subsection{Droplet spreading}
\label{sec:granuspreading}

The addition of particles to a liquid phase that spreads on a solid substrate introduces a new length scale as well as a new source of dissipation which are not accounted for in the description provided in \S\,\ref{sec:drop-spreading}.
Moreover, the simplistic effective medium approach of granular suspensions is challenged, as particle confinement increases as spreading proceeds.
These considerations suggest that the spreading of a droplet of suspension may be quite complex.

Surprisingly, the spreading dynamics of granular suspensions are qualitatively not different from that of a simple liquid.
The dynamic contact angle depends on the contact line velocity following a Cox-Voinov-like law \citep{zhao2020spreading,pelosse2023probing}, as shown in Fig.\,\ref{fig:CV_Tanner_susp}(a), and the radius of large droplets grows following the gravity-viscous power law, $R(t)\propto t^{1/8}$ \citep{pelosse2024success}, as seen in Fig.\,\ref{fig:CV_Tanner_susp}(b).
However, the quantity  $(\eta_{\rm eff}-\eta_0)/(\eta_\mathrm{bulk}-\eta_0)$, which goes to 0 as $\eta_\mathrm{eff}\rightarrow \eta_0$ and 1 as $\eta_\mathrm{eff}\rightarrow \eta_\mathrm{bulk}$, indicates that for both laws, the effective viscosity of the suspension during spreading, $\eta_\mathrm{eff}$, is smaller than the bulk value, $\eta_\mathrm{bulk}$ \citep{zhao2020spreading}, see Fig.\,\ref{fig:CV_Tanner_susp}(c).
\begin{figure*}[!ht]
    \centering\includegraphics[width = \linewidth]{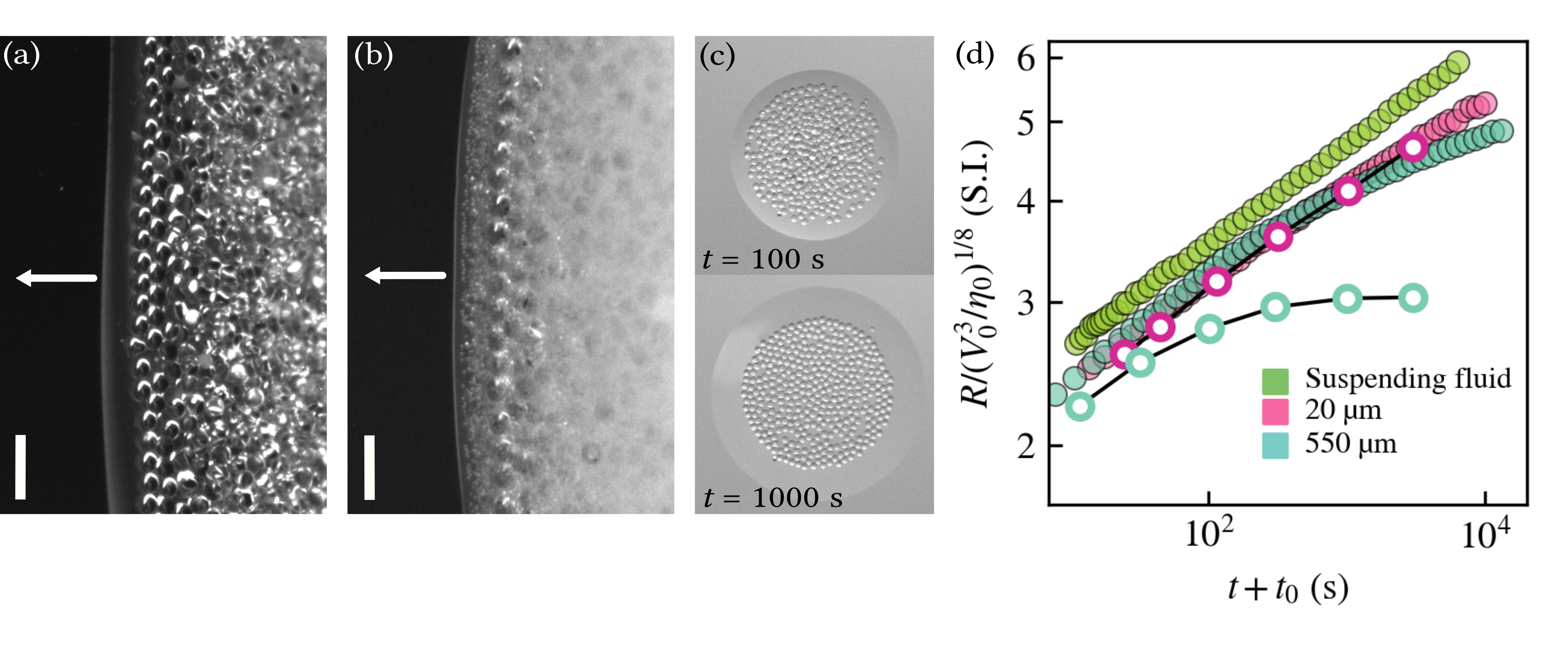}
    \caption{Top-view of an advancing contact line for a droplet made of (a) a monomodal suspension with 80-\si{\micro\meter} particles and (b) a bimodal suspension with 10-\si{\micro\meter} and 80-\si{\micro\meter} particles. Scale bar: \SI{200}{\micro\meter}. (c) Top view snapshots at $t =$\,\SI{100}{\second} (top) and  $t =$\,\SI{1000}{\second} (bottom), for a drop made with \SI{550}{\micro\meter} particles with $\phi = 40\%$ and $V_0=$\SI{114}{\micro\liter}.
    (d) Evolution of the liquid outer radius (solid symbols) and particle-phase outer radius (open symbols), normalized by volume and viscosity effect for drops made of pure liquid and granular suspensions at $\phi=40\%$ (see legend).
    All adapted from \citep{pelosse2023ecoulements}.}
    \label{fig:top-view-CL}
\end{figure*}
Moreover, when tracking radius growth, measurements indicate that $\eta_{\rm eff}$ has a magnitude between 20 and 25\% of the bulk value and is independent of particle size $d$, similarly to $\eta_\mathrm{bulk}$.
However, in the relation between the apparent contact angle and velocity, $\eta_{\rm eff}$ becomes a decreasing function of $d$ for $d\leq 100$\,\si{\micro\metre}, and becomes equal to the viscosity of the suspending liquid for larger particle sizes. 
The latter finding results from the fact that particles can populate the viscous-capillary region and approach the vicinity of the contact line so long as the liquid/gas interface height $h$ is larger than the particle diameter $d$.
This results in either a particle-depleted area when the particles are monodisperse, shown in Fig.\,\ref{fig:top-view-CL}(a), or a region rich in the smaller particles for a bi-disperse suspension, see Fig.\,\ref{fig:top-view-CL}(b), in which case the effective viscosity of the suspension is selected by the ability of these small particles to flow through the channels of the porous matrix formed by the large ones and enter the viscous-capillary corner.

Another striking feature on these top views is that the first few rows closest to the contact line display ordering and even crystallization, see Fig.\,\ref{fig:top-view-CL}(a).
This confinement-induced ordering is consistent with the discussion in \S\,\ref{sec:thinFilmFlows}, and the ordering decays over a few particle diameters as the confinement of the particulate phase weakens.
Particles that enter this region remain trapped and only slow rearrangements occur as the perimeter of the droplet increases.
Ordered particles, trapped by the free interface, likely slide on the surface of the substrate.
Finally, the particles and the suspending liquid do not spread at the same velocity. 
The distance between the particle-rich region and the contact line increases over time, as the height of the liquid/gas interface decreases, see Fig.\,\ref{fig:top-view-CL}(c).
For large particles, $d=550$ \si{\micro\metre}, separation occurs early in the process: after \SI{100}{\second} of spreading, a particle-rich phase occupies the central region of the droplet, surrounded by a particle-depleted volume, evidenced in Fig.\,\ref{fig:top-view-CL}(c).
From this point on, the particle phase spreads only marginally, while the liquid contact line continues to advance. 
Both the arrest radius and the arrest time of the particle phase decrease with increasing particle size.
The two phases nevertheless remain coupled, as evidenced by the reduction in the growth rate of the liquid radius once the particle phase stops spreading, see Fig.,\ref{fig:top-view-CL}(d). 
This transition occurs when the interface height becomes comparable to the particle diameter, and leads to a new regime in which the viscous-gravity power-law growth of the radius no longer applies. 
Then, the contact line advances under the supplementary constraint that the liquid trapped in the particle phase has an increased resistance to flow.
The spreading dynamics become similar to that of a meniscus rising to its equilibrium height on vertical plate \citep{clanet2002a}.
This transition from a Newtonian-like to a heterogeneous flow regime when particle confinement increases is similar to that reported in other capillary flows of granular suspensions, such as dip-coating \citep{gans2019dip, palma2019dip, sauret2019capillary} and droplet pinch-off \citep{bonnoit2012accelerated, mathues2015capillary, zhao2015inhomogeneity, chateau2018pinch, thievenaz2022onset}.

\begin{figure*}[!htb]
    \centering
    \includegraphics[width=0.49\linewidth]{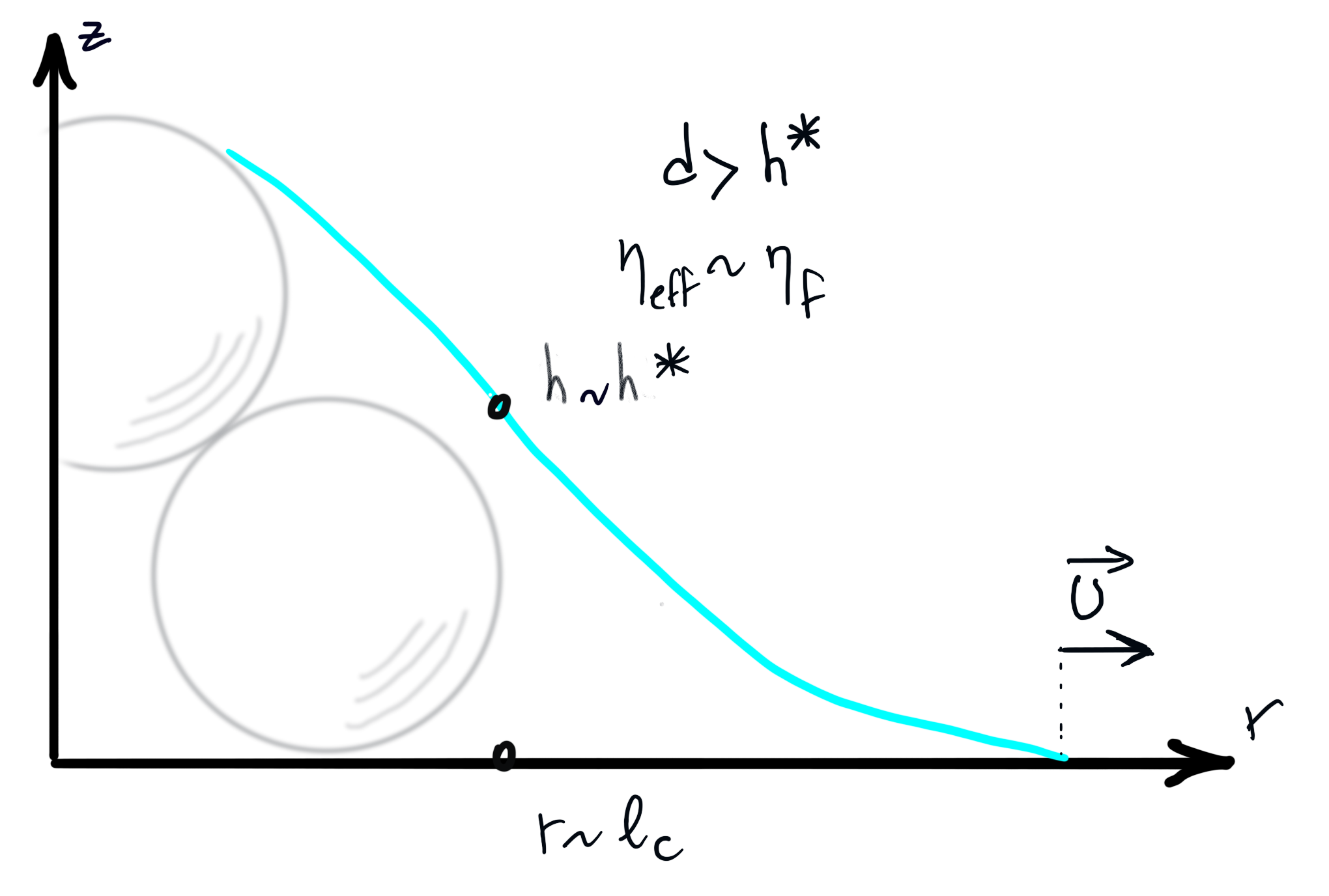}
    \includegraphics[width=0.49\linewidth]{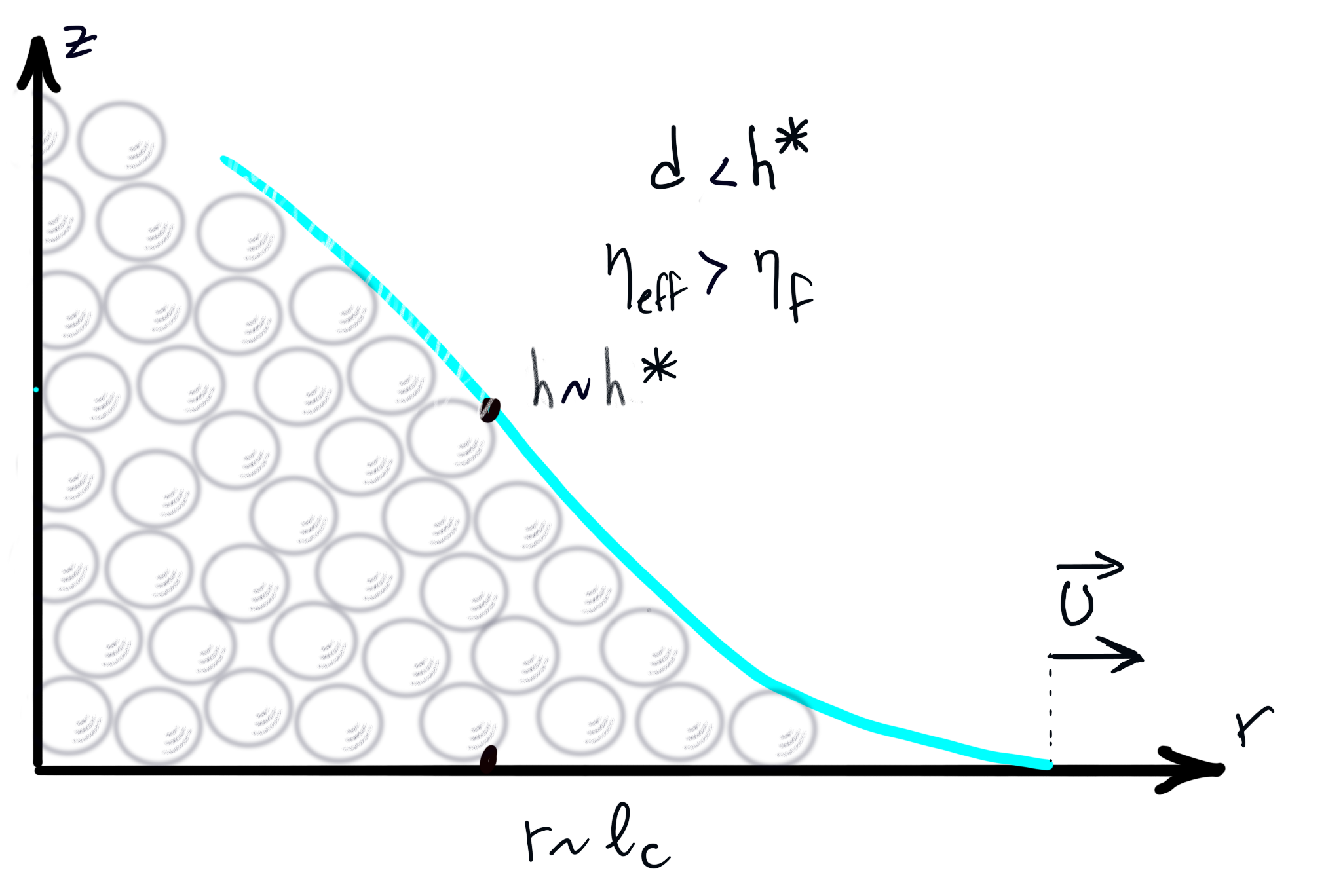}
    \caption{Granular particles next to an advancing contact line. 
    The effective wetting viscosity of the suspension is that of the pure fluid if $d>h^\star$, or larger if particles can enter the dissipative region when $d<h^\star$.}
    \label{fig:fig5}
\end{figure*} 

Thus, while their spreading dynamics may appear to be simple at first glance, suspensions display a micro-structural complexity associated to an unexpectedly low value of their effective viscosity that challenges our understanding.
We provide below a quick recapitulation of what is understood and what is not at the time of writing.
Let us focus on monodisperse suspensions.
The results of section \ref{sec:drop-profile} show that, for simple Newtonian liquids, the viscous-capillary region extends up to a distance $\ell_c$ from the contact line, where the height of the liquid/gas interface is $h^\star \sim \SI{100}{\micro\meter}$ over the range of capillary numbers that we investigate, given by Eq.\,\eqref{eq:spreadingf-h-star}. 
This height corresponds to the particle size beyond which the effective viscosity of the suspension is equal to that of the suspending liquid.
Particles then contribute to dissipation when entering the viscous-capillary region, i.e. when $d \lesssim h^\star$, while spreading is similar to that of the pure fluid when $d > h^\star$, i.e., when particles cannot access the viscous–capillary region \citep{pelosse2023probing}, as sketched in Fig.\,\ref{fig:fig5}.

These considerations rationalize the dependence of $\eta_\text{eff}$ on $\phi$ and $d$, unlike $\eta_\text{bulk}$ that only depends on $\phi$. 
The situation is much less clear in the viscous-gravity regime.
Maybe the most puzzling fact is that the apparent viscosity is independent of the particle size.
The velocity difference between the contact line and the edge of the particle-rich region suggests that a contribution related to the drainage of this latter porous region could play a role.
Then, an argument related to the effective viscosity of ordered porous media \citep{brinkman1949} may be used to estimate the viscosity \citep{pelosse2024success}.
However, while this argument sounds reasonable for large particles, it is rather fragile when considering small particles, as the velocity difference decreases.
Moreover, the main contribution of this mechanism is apparently to slow down radius growth rather than change the viscosity.
Thus, the physics at play to explain the lower apparent viscosity compared to the bulk measured at the scale of the droplet remains to be identified.

In conclusion, the spreading of a dense granular suspension is similar to that of a simple Newtonian fluid, albeit faster than expected from bulk properties.
The lower value of their effective viscosity both in the viscous-capillary and in the gravity-viscous regime, compared to the bulk stems from several effects induced by confinement, such as particle depletion and ordering.
Long-time dynamics is also interesting, as spreading slows down.
The suspending liquid forms a structure similar to a meniscus that relaxes slowly to an equilibrium position that must accommodate the connection of the contact line to an immobile porous particle-rich region near the center of the droplet.

\subsection{Dip-coating of granular suspensions}
\label{sec:granudipcoating}

\begin{figure*}[!htb]
    \centering
    \includegraphics[scale=1]{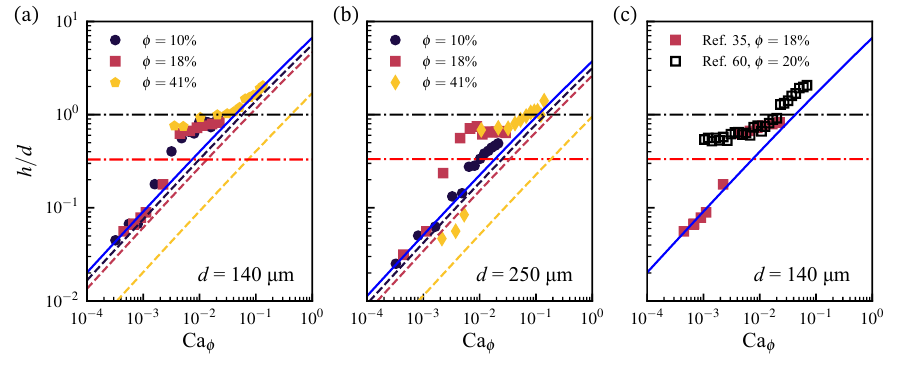}
    \caption{Summary of the experimental data available concerning the dip-coating of solid plates with granular suspensions. Thickness $h$ of the entrained film in units of particle diameter $d$ as a function of the suspension capillary number $\text{Ca}_{\phi}=\eta(\phi)U/\gamma$ for a suspension of (a) 140--\si{\micro\metre} and (b) 250-\si{\micro\metre} particles in silicone oil, extracted from \citet{gans2019dip}. 
    (c) Comparison of two datasets obtained for  140-\si{\micro\metre} \citep{gans2019dip} and 145-\si{\micro\metre} \citep{palma2019dip} particles at similar volume fractions.
    In all panels, the continuous blue line is the White-Tallmadge generalization of the Landau-Levich-Derjaguin law \eqref{eq:WT} calculated with $\text{Ca}_{\phi}$. 
    The horizontal dashed lines represent the same law for the suspending liquid alone. 
    The dot-dashed lines represent $h/d=1$ (black) and $h/d=1/3$ (red).
    Symbol colors encode the volume fraction of the suspension following the legend. 
    The shift of the suspending-liquid curves reflects the value of the viscosity ratio coefficient $\eta(\phi)/\eta_0$. }
    \label{fig:dipCoatExpSusp}
\end{figure*}
Coming back to the dip-coating problem and considering granular suspensions, the obvious questions are whether the classical, continuum approach can be applied, and what are the conditions under which particles will be entrained in the film.
%
%
These issues were first addressed by \citet{colosqui2013b}, following reports that showed that the withdrawal of a plate from a suspension led to a self-assembled stripe-pattern \citep{ghosh2007,watanabe2009}.

\citet{colosqui2013b} demonstrated that the streamline originating from the stagnation point divides the flow in the liquid between a sheared region close to the wall and a recirculating volume away from it.
The thickness of the sheared region is comparable to the height $h^\star_\text{LLD}$ of the stagnation point given by Eq.\,\eqref{eq:stagPos}.
\citet{colosqui2013b} showed with numerical simulations that the condition for the entry of a single particle into the film is $h^\star_\text{LLD}>d$.
Assuming that $h_0/\ell_c<<\sqrt{\text{Ca}}$, Eq.\,\eqref{eq:stagPos} leads to: 
\begin{equation}
    h^{\star}_\text{LLD} \simeq 3h_0 > d \quad\Rightarrow\quad
    h_0 > \frac{d}{3}.
    \label{eq:EntryCond}
\end{equation}
Rather counter-intuitively, a single particle can therefore enter a film having a thickness equal to a third of its diameter.
Introducing the particle Bond number $\text{Bo}_\text{p}=(d/\ell_c)^2$, and replacing $h_0$ with Eq.\,\eqref{eq:LLDlaw}, Eq.\,\eqref{eq:EntryCond} leads to:
\begin{equation}
    \text{Ca} > 0.209~\text{Bo}_\text{p}^{3/4}.
    \label{eq:nonDimEntryCond}
\end{equation}

When $h^\star_\text{LLD}/d\leq 1$, a particle trapped in the sheared region rises to a steady position in the meniscus. 
This scenario may lead to a closely-packed assembly of particles in the meniscus.
Because the meniscus flattens while the drag on the assembly exceeds that on a single particle, particle entry occurs at lower values of $h^\star_\text{LLD}/d$.

These theoretical findings were discussed later in two experimental articles \citep{gans2019dip,palma2019dip}. 
Figure \ref{fig:dipCoatExpSusp}(a) and (b) shows results obtained for suspensions of 140-\si{\micro\metre} and 250-\si{\micro\metre} polystyrene particles suspended in a density-matched silicone oil \citep{gans2019dip}.
The data are compared to four different versions of the White-Tallmadge (WT) prediction \eqref{eq:WT}.
The first one uses the suspension capillary number computed from the suspension bulk viscosity $\text{Ca}_{\phi}=\eta(\phi)U/\gamma$. 
The other three are obtained by dividing $\text{Ca}_{\phi}$ by $\eta(\phi)/\eta_0$, i.e. using the capillary numbers for the suspending liquid alone, and testifying of the shift in viscosity due to particle addition.
For both particle sizes, film thickness grows in qualitative agreement with Eq.\,\eqref{eq:WT} at capillary numbers $\text{Ca} < 2\times 10^{-3}$, whatever the volume fraction.
The authors report that the entrained films are particle-free.
Hence, the thickness of the thin film should follow the WT law for the suspending liquid, i.e., the dashed downward-shifted curves.
However, this hypothesis turns out to be wrong. 
The 10\% and 18\% datasets collapse onto one another and are well described by the WT prediction using the suspension bulk viscosity (blue solid line).
The 41\% 250-\si{\micro\metre} dataset lies below the other ones on Fig.\,\ref{fig:dipCoatExpSusp}(b), between the WT laws for the suspending liquid and for the suspension.
In fact, a fit employing an effective viscosity, analogous to the approach taken for spreading, yields a value about 25\% of the bulk viscosity, i.e., a result comparable to the wetting studies described earlier in \S\,\ref{sec:granuspreading}.

Increasing further the capillary number triggers particle entrainment, for thicknesses slightly smaller than the single-particle threshold \eqref{eq:EntryCond}.
This is likely due to collective effects that induce meniscus flattening and increased drag \citep{colosqui2013b}.
The film thickness then becomes a slowly increasing function of the capillary number.
Beyond $h/d_p\simeq1$, the thickness increases faster again, following Eq.\,\eqref{eq:WT} evaluated on $\text{Ca}_{\phi}$, as expected.

Figure \ref{fig:dipCoatExpSusp}(c) compares datasets obtained for suspensions of particles of almost identical size, 140\,\si{\micro\metre} and 145\,\si{\micro\metre}, dispersed at volume fractions of 18 and 20\%.
Surprisingly, the two datasets do not collapse perfectly.
Differences in substrate wetting may explain this discrepancy. 
Indeed, \citet{gans2019dip} use a system exhibiting complete wetting.
They confirmed that their suspending liquids follow the LLD law when coating their substrates, as seen in Fig.\,\ref{fig:summaryLLD}.
By contrast, \citet{palma2019dip} report finite contact angles, indicating only partial wetting of the substrate.
Then, dynamics akin to those described in subsection \ref{subsec:partWet} may play a role.

\citet{palma2019dip} dismissed this explanation because they observed very thin films even in the particle-free regime.
However, the thicknesses they report, of the order of several hundreds \si{\nano\meter}, are incompatible with the at most \SI{10}{\micro\meter} predicted by the WT (or LLD even) law over this range of capillary numbers.
\begin{figure*}
    (a)\hspace{5cm}(b)\hspace{5cm}(c)\hspace{5cm} \\
    \centering    \includegraphics[width=0.99\linewidth]{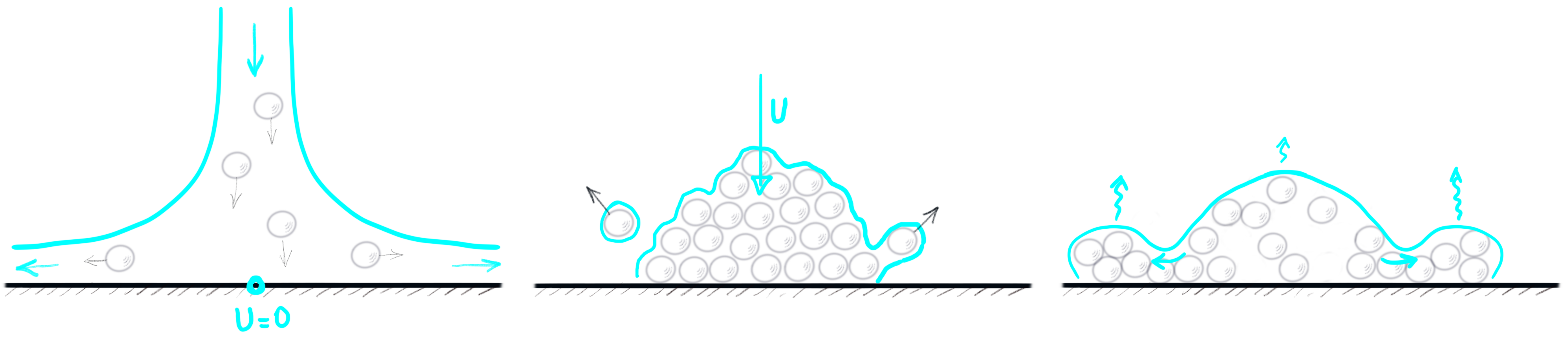}
    \caption{(a) Jet of liquid containing particles used for polishing or abrasion, (b) splashing of a drop containing particles impacting on a solid surface, and (c) sessile drop containing particles, with accumulation of solid next to the edge due to the enhanced liquid evaporation and particle confinement by the free interface.}
    \label{fig:conclusion}
\end{figure*}

Therefore, it is hypothesized that the observed withdrawal transition is qualitatively similar to the one described by \citet{snoeijer2006c}, albeit at lower values of the critical capillary number $\text{Ca}_c$.
The decrease of $\text{Ca}_c$ may arise because, under partial wetting conditions, film entrainment is a subcritical transition that can be triggered by contact-line perturbations, such as the presence of particles. 
At larger capillary numbers, \citet{palma2019dip}'s data lie well above the WT law, despite following a qualitatively similar tendency.
The thickness of the film seems influenced by an additional yet unknown contribution.
Finally, their $\phi=50\%$ dataset suggests that the film thickness selected in the monolayer regime depends on the volume fraction, and may exceed one particle diameter.
The structure of this layer would be interesting to study, as it is not strictly a monolayer anymore.

These puzzling observations underscore the necessity for further research, which will be of interest to both the suspension and the wetting community.

\section{Conclusion and perspectives}
\label{sec:conclusion}

In conclusion, the behaviors of thin-film flows of granular suspensions near solid boundaries reveals the dual nature of these systems. 
When the particles are small compared to the film thickness, the suspension can be treated as a continuous medium with effective properties. 
However, under stronger confinement, the discrete character of the suspension emerges, requiring a two-phase description with distinct behaviors.

This complex coupling between the free surface and the suspension flow arises in situations such as drop spreading and dip coating.
In both cases, the particles locally enhance dissipation and modify the free-surface flow where viscous effects dominate, while the interfacial confinement also acts on the particle phase, particularly its microstructure.

In this review, we limited our focus to the situation of capillary flow of granular suspensions (micron-scale particles and larger) next to a solid wall at low Reynolds, i.e., low inertia and high flow dissipation. 
Accounting for inertia however, becomes necessary in situations such as suspension jets and drop impact.
In the first situation, sketched in Fig.\,\ref{fig:conclusion}(a), relevant for abrasive/polishing jet technologies, the trajectories and speed of the impinging particles set the abrasion efficiency.
Due to inertia, particle speed may differ from that of the fluid streamlines, in particular near the stagnation point of the jet (where fluid speed $U$ is equal to zero) , but also in the surrounding fluid film where particles are subject to high shear stresses \citep{buss2022towards}.
In drop impact, pictured in Fig.\,\ref{fig:conclusion}(b), the additions of particles varying impact speed, fluid viscosity, particle size, and volume fraction can lead to spreading, splashing (or break-up), or even bouncing \citep{nicolas2005spreading, marston2013impact, peters2013splashing, shah2024drop}. 
The bouncing occurs in very dense suspensions, in a region of the parameter space where pure liquid droplets cannot bounce.
This acquired bouncing ability of the drops would come from an effective solid-like elasticity of the particle packing, jammed by the free interface confinement, and distorted upon impact \citep{klein2016splashing}. 

The scope of the review was further constrained to systems characterized by constant fluid and particle volumes.
However, relaxing this constraint, for instance by introducing fluid evaporation, is pertinent in many situations and adds another layer of complexity, potentially leading to particle accumulation near the interface, shown in Fig.\,\ref{fig:conclusion}(c).
A familiar example is the formation of coffee stains, where particle-dense rings are left behind after droplet evaporation. 
This phenomenon is relevant in printing technologies, biological phenomena, thin-film deposition, and self-assembly techniques.
It is typically observed for sessile droplets with a pinned contact line containing particles up to micron size, and can be explained by the accumulation of particles at the rim of the drop due to enhanced evaporation in this region \citep{deegan1997capillary}.
Due to the importance of capillary interactions between particles and the interface, this effect can be suppressed by using particles of different shapes \citep{yunker2011suppression}. 
On hydrophobic substrates, where the contact line recedes and droplets remain more spherical, particle accumulation near the interface leads to a qualitatively different situation: a shell of particles grows and rearrange into an extremely dense and crystalline packing at the interface as evaporation proceeds \citep{bamboriya2023universality}. 
For sufficiently small (but still above the micron scale) and soft particles, this shell can undergo buckling, a phenomenon not observed for larger, stiffer, or high-aspect-ratio particles \citep{al2020increasing}.

These are just a few examples of avenues beyond viscous flow and constant fluid and particle volumes that come to mind.  
Undoubtedly, there are additional complexities in the behavior of thin film suspensions that remain to be explored, such as the dynamics of moving suspension droplets, the onset and evolution of instabilities within thin suspension films, and related phenomena to the coupling between particle interactions and thin film flow.



\end{document}